\documentclass[twoside,twocolumn]{article}
\usepackage[super,sort&compress,comma]{natbib} 
\usepackage{times,mathptmx}
\usepackage{sectsty}
\usepackage{balance} 
\usepackage{amssymb}
\usepackage{graphicx} 
\usepackage{lastpage}
\usepackage[format=plain,singlelinecheck=false,font=small,labelfont=bf,labelsep=space]{caption} 
\usepackage{fancyhdr}
\usepackage{color}
\usepackage{calrsfs}
\usepackage{siunitx}
\DeclareMathAlphabet{\pazocal}{OMS}{zplm}{m}{n}

\usepackage{url,hyperref}
\usepackage[usenames,dvipsnames]{xcolor}
\hypersetup{colorlinks=true, linkcolor=BrickRed, urlcolor=blue!50!black, citecolor=blue!50!black}

\begin{document}

\newcommand{\red}{\textcolor{red}}
\newcommand{\green}{\textcolor{green}}
\newcommand{\blue}{\textcolor{blue}}

\thispagestyle{plain}
\fancypagestyle{plain}{
\fancyhead[L]{\includegraphics[height=8pt]{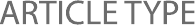}}
\fancyhead[C]{\hspace{-1cm}\includegraphics[height=20pt]{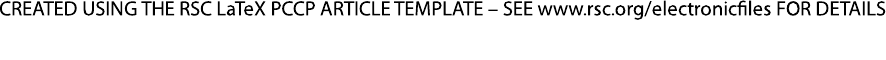}}
\fancyhead[R]{\includegraphics[height=10pt]{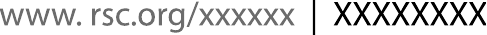}\vspace{-0.2cm}}
\renewcommand{\headrulewidth}{1pt}}
\renewcommand{\thefootnote}{\fnsymbol{footnote}}
\renewcommand\footnoterule{\vspace*{1pt}%
\hrule width 3.4in height 0.4pt \vspace*{5pt}} 
\setcounter{secnumdepth}{5}

\makeatletter 
\def\subsubsection{\@startsection{subsubsection}{3}{10pt}{-1.25ex plus -1ex minus -.1ex}{0ex plus 0ex}{\normalsize\bf}} 
\def\paragraph{\@startsection{paragraph}{4}{10pt}{-1.25ex plus -1ex minus -.1ex}{0ex plus 0ex}{\normalsize\textit}} 
\renewcommand\@biblabel[1]{#1}            
\renewcommand\@makefntext[1]%
{\noindent\makebox[0pt][r]{\@thefnmark\,}#1}
\makeatother 
\renewcommand{\figurename}{\small{Fig.}~}
\sectionfont{\large}
\subsectionfont{\normalsize} 

\fancyfoot{}
\fancyfoot[LO,RE]{\vspace{-7pt}\includegraphics[height=9pt]{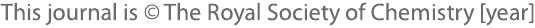}}
\fancyfoot[CO]{\vspace{-7.2pt}\hspace{12.2cm}\includegraphics{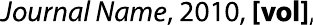}}
\fancyfoot[CE]{\vspace{-7.5pt}\hspace{-13.5cm}\includegraphics{headers/RF}}
\fancyfoot[RO]{\footnotesize{\sffamily{1--\pageref{LastPage} ~\textbar  \hspace{2pt}\thepage}}}
\fancyfoot[LE]{\footnotesize{\sffamily{\thepage~\textbar\hspace{3.45cm} 1--\pageref{LastPage}}}}
\fancyhead{}
\renewcommand{\headrulewidth}{1pt} 
\renewcommand{\footrulewidth}{1pt}
\setlength{\arrayrulewidth}{1pt}
\setlength{\columnsep}{6.5mm}
\setlength\bibsep{1pt}

\twocolumn[
  \begin{@twocolumnfalse}
\noindent\LARGE{\textbf{Characterizing anomalous diffusion in crowded polymer solutions and gels over five decades in time with variable-lengthscale fluorescence correlation spectroscopy}}
\vspace{0.6cm}

\noindent\large{\textbf{Daniel S. Banks,\textit{$^{a\dag}$} Charmaine Tressler,\textit{$^{a}$} Robert D. Peters,\textit{$^{a}$} Felix H\"ofling,\textit{$^{bcd}$} and
C\'ecile Fradin$^{\ast}$\textit{$^{ae}$}}}\vspace{0.5cm}

\noindent\textit{\small{\textbf{Received 20th May 2015, Accepted 27th March 2016\newline
}}}

\noindent \textbf{\small{DOI: 10.1039/c5sm01213a}}
\vspace{0.6cm}

\noindent \normalsize{The diffusion of macromolecules in cells and in complex fluids is often found to deviate from simple Fickian diffusion. One explanation offered for this behavior is that molecular crowding renders diffusion anomalous, where the mean-squared displacement of the particles scales as $\langle r^2 \rangle \propto t^{\alpha}$ with $\alpha < 1$. Unfortunately, methods such as fluorescence correlation spectroscopy (FCS) or fluorescence recovery after photobleaching (FRAP) probe diffusion only over a narrow range of lengthscales and cannot directly test the dependence of the mean-squared displacement (MSD) on time. Here we show that variable-lengthscale FCS (VLS-FCS), where the volume of observation is varied over several orders of magnitude, combined with a numerical inversion procedure of the correlation data, allows retrieving the MSD for up to five decades in time, bridging the gap between diffusion experiments performed at different lengthscales. In addition, we show that VLS-FCS provides a way to assess whether the propagator associated with the diffusion is Gaussian or non-Gaussian. We used VLS-FCS to investigate two systems where anomalous diffusion had been previously reported. In the case of dense cross-linked agarose gels, the measured MSD confirmed that the diffusion of small beads  was anomalous at short lengthscales, with a cross-over to simple diffusion around $\approx 1~\mu$m, consistent with a caged diffusion process. On the other hand, for solutions crowded with marginally entangled dextran molecules, we uncovered an apparent discrepancy between the MSD, found to be linear, and the propagators at short lengthscales, found to be non-Gaussian. These contradicting features call to mind the ``anomalous, yet Brownian'' diffusion observed in several biological systems, and the recently proposed ``diffusing diffusivity'' model. }
\vspace{0.5cm}
 \end{@twocolumnfalse}
  ]

\footnotetext{\textit{$^{a}$~Department of Physics and Astronomy, McMaster University, 1280 Main St. W, Hamilton, ON L8S 4M1, Canada.}}
\footnotetext{\textit{$^{b}$~Max-Planck-Institut f\"ur Intelligente Systeme, Heisenbergstr. 3, 70569 Stuttgart, Germany}}
\footnotetext{\textit{$^{c}$~IV. Institut f\"ur Theoretische Physik, Universit\"at Stuttgart, Pfaffenwaldring 57, 70569 Stuttgart, Germany.}}
\footnotetext{\textit{$^{d}$~Fachbereich Mathematik und Informatik, Freie Universit\"at Berlin, Arnimallee 6, 14195 Berlin, Germany.}}
\footnotetext{\textit{$^{e}$~Department of Biochemistry and Biomedical Sciences, McMaster University, 1280 Main St. W, Hamilton, ON L8N 3Z5, Canada. E-mail: fradin@physics.mcmaster.ca}}
\footnotetext{\dag~Present address: Chalk River Laboratories, Chalk River, ON K0J 1J0, Canada.}

\section{Introduction}

Diffusion is a ubiquitous process observed in very different contexts. In most systems, diffusion is ``simple'', or ``Fickian'', \textit{i.e.} characterized by a constant diffusion coefficient inversely proportional to the medium viscosity, by a Gaussian distribution of displacements, and - the hallmark of simple diffusion - by a mean-squared displacement that is directly proportional to time. In biological systems, however, and in particular in cells, diffusive motions are often found to deviate from simple Fickian diffusion \cite{Klafter2005}. This deviation can take different forms, from an apparent size-dependent viscosity of the medium to a non-Gaussian distribution of displacements or to a mean-squared displacement that is not directly proportional to time\cite{Luby1987,Kusumi1993,Feder1996,Wachsmuth2000,Platani2002,Weiss2003,Wawrezinieck2005,Golding2006,Guigas2007,Bancaud2009,Abu2010, Jeon2011}. Similar effects can be recreated \textit{in vitro}, using for example polymer solutions \cite{Busch2000,Weiss2004,Banks2005}, gels \cite{Amblard1996,Valentine2001,Wong2004,FatinRouge2004} or colloidal suspensions \cite{Weeks2002}. This suggests that ``anomalous'' diffusion, rather than being a property of active matter, is a characteristic of complex fluids.

Many possible reasons have been advanced to explain deviations from simple Fickian diffusion in complex environments, notably temporary confinement \cite{Kusumi1993,Weeks2002,Lenne2006}, molecular crowding \cite{Saxton1994,Weiss2004,Banks2005,Szymanski2009} and reversible binding to traps \cite{Saxton1996,Bronstein2009}.
Accordingly, numerous models have been proposed to describe anomalous diffusion processes (see Refs. \citenum{Metzler2000,Sokolov2012,Hofling2013,Bressloff2013,Metzler2014,Weiss2014,Meroz2015} for reviews). This term refers to a diffusive motion for which the mean-squared displacement of the tracer particles scales as a power law in time, $r^2(t) \propto t^\alpha $, with an anomalous exponent $\alpha  \neq 1$. If $\alpha <1$, the process is referred to as sub-diffusion. Most realistic models of diffusion in complex media predict that the anomalous behaviour will occur only over a certain range of lengthscales (prescribed by characteristic lengthscales in the system), with a mean-squared displacement  linear in time outside that range \cite{Saxton1994,Weeks2002,Yeung2007}. 

Experimentally distinguishing between different diffusion models is challenging since each technique has its limitations and tends to probe diffusion only at a single lengthscale. One exception in that regard is single particle tracking, which has the potential to directly test the lengthscale dependence of diffusion through direct measurements of the mean-squared displacement. It has successfully been used to characterize non-Fickian diffusion in a number of systems \cite{Amblard1996,Weeks2002,Fujiwara2002}. However, in its simplest form, single particle tracking does not have the time-resolution necessary to study the fast three-dimensional diffusion of biomolecules in aqueous environments. Thus it cannot be used to investigate short-timescale diffusion in the cell interior or in crowded solutions. Fluorescence recovery after photobleaching suffers from the same limitation, with the timescale of the processes that can be investigated limited by the duration of the photobleaching step \cite{Waharte2005}. In contrast, fluorescence correlation spectroscopy (FCS) has a time-resolution that is only limited by the photophysics of the fluorophores used to label the tracer particles and can easily reach $10~\mu$s.

FCS traditionally provides information on diffusion around a single lengthscale, fixed by the radius ($w_0$) of the usually diffraction-limited confocal observation volume. It gives access to the characteristic time taken by fluorophores to diffuse through this observation volume ($\tau_D$). This, in turn, yields the particle diffusion coefficient, $D = w_0^2/\left( 4 \tau_D \right)$. A more precise analysis of the correlation data also allows detecting deviations from simple Fickian diffusion, and extracting a value for the anomalous exponent of the diffusion at this particular lengthscale \cite{Schwille1999,Wu2008}. Recently, it has been pointed out that, in the case of a Gaussian distribution of displacements, an inversion procedure could be used to extract from the ACF the mean-squared displacement over several decades in time around $\tau_D$ \cite{Shusterman2004,Shusterman2008,Horton2010,Hofling2013}. In parallel, the use of variable-lengthscale FCS (VLS-FCS), where the value of $w_0$ is systematically varied, has been introduced to further probe the lengthscale dependence of diffusion \cite{Wawrezinieck2004,Wawrezinieck2005,Masuda2005}. Its potential to test the ``diffusion law'', or dependence of the mean-squared displacement ($ \propto w_0^2$) on time ($\tau_D$), has been tapped into to study diffusion in membranes \cite{Wawrezinieck2004,Wawrezinieck2005,Eggeling2008,Hofling2011,Favard2011} and in polymer solutions \cite{Masuda2005,Masuda2005b,Masuda2006,King2014}.

In a previous study, we found using conventional single-scale confocal FCS that the diffusion of proteins in aqueous solutions crowded by large molecular weight dextrans was consistent with anomalous diffusion \cite{Banks2005}. Other studies in similar systems have also suggested that protein diffusion in crowded marginally entangled polymer solutions deviates from simple Fickian diffusion \cite{Weiss2004,Sanabria2007,Goins2008,Szymanski2009,Pan2009,Ernst2012}. This, however, remains somewhat controversial, as mobile obstacles are not expected to produce anomalous diffusion \cite{Saxton1994}. Further, experiments using either a different tracer or a different polymer as a crowder showed opposite results \cite{Dauty2004,Dix2008,Shakhov2012}. In this study, we used VLS-FCS combined with the inversion procedure of the ACF mentioned above to test the time-dependence of the mean-squared displacement in two different ways. This allowed revisiting the nature of the diffusion of tracer proteins in marginally entangled polymer solutions, and comparing it with that of beads in agarose gels, a model system for cross-linked polymer networks that has been unambiguously shown to support anomalous diffusion \cite{Valentine2001,Lead2003,FatinRouge2004,FatinRouge2006,Labille2007}.

\section{Theory} \label{theory}

In this section, we briefly discuss the derivation of the autocorrelation function (ACF) obtained as the output of an FCS experiment in the general case of a diffusion process with a Gaussian distribution of displacements. We then apply this derivation to two particular cases, simple diffusion and fractional Brownian motion.

\subsection{Diffusion processes with a Gaussian propagator} \label{inversion}

The ACF obtained as the result of an FCS experiment is defined as $G(\tau)= \langle \delta I(t) \, \delta I(t+\tau) \rangle/\langle I(t) \rangle^2$, where $I(t)$ is the collected fluorescence signal, brackets denote a time average, and $\delta I (t) = I(t)-\langle I \rangle$. For a diffusion process, the ACF can be calculated from the isotropic propagator, $p(\rho,\tau)$ (also known as distribution of displacements, as it gives the probability for the displacement of a particle to be equal to $\rho$ after a time $\tau$):
\begin{equation}
G \left( \tau \right) = \frac{1}{ \left \langle c \right \rangle} \frac{ \int d{\vec r}~\pazocal{O} \left( {\vec r} \right)   \int d{\vec r'} ~\pazocal{O} \left( {\vec r'} \right)   p \left(\left| {\vec r} - {\vec r'} \right| ,\tau \right)}
{  \left( \int d{\vec r}~ \pazocal{O} \left( {\vec r} \right)   \right)^2 }
\label{ACF1}
\end{equation}
This expression can be calculated in the commonly considered case of a three-dimensional (3D) Gaussian observation volume (characterized by its $1/\mathrm{e}^2$ radius, $w_0$, and its aspect ratio, $S$), $ \pazocal{O} \left( {\vec r} \right) = \exp(-2x^2/ w_0^2 -2y^2/ w_0^2 -2z^2/ (Sw_0)^2)$, and a 3D Gaussian propagator:
\begin{equation}
p \left( \rho,\tau \right) =\left( \frac{2 \pi}{3} MSD \left( \tau \right) \right)^{-\frac{3}{2}} \exp\left(-\frac{3\rho^2}{2\, MSD(\tau)}\right)
\label{propagator}
\end{equation}
where $MSD(\tau)$ is the mean-squared displacement of the particles. In that case, the ACF takes a simple form\cite{Shusterman2004,Hofling2013}:
\begin{equation}
G \left( \tau, w_0 \right) = \frac{1} {N} \left[ 1 + \frac{MSD \left( \tau \right) }{3 w_0^2/2} \right]^{-1} \left[ 1 + \frac{MSD \left( \tau \right) }{3 (Sw_0)^2/2} \right] ^{-\frac{1}{2}},
\label{Eq:ACFG}
\end{equation}
where $N$ is is the average number of fluorescent particles found in the integrated observation volume, $V = S w_0^3 \pi^{3/2}$. In Eq.~\ref{Eq:ACFG} the dependence on $w_0$ was made explicit, to help the interpretation of VLS-FCS experiments. If the value of $w_0$ is known, then Eq.~\ref{Eq:ACFG} can be numerically inverted for each lag time in order to obtain the mean-squared displacement as a function of time. It is important to keep in mind, however, that this procedure will return the actual mean-squared displacement only if the considered diffusion process has a Gaussian and isotropic propagator.

\subsection{Simple Fickian diffusion}

The simplest example of a diffusion process with a Gaussian propagator (Eq. \ref{propagator}) is simple Fickian diffusion, with a mean-squared displacement linear in time, $MSD(t) = 6Dt$.
The diffusion coefficient is given by the Stokes-Einstein relationship, $D = kT / \left( 6 \pi \eta R \right)$, where $R$ is the hydrodynamic radius of the particle, $\eta$ is the fluid viscosity, $k$ is Boltzmann's constant and $T$ is the absolute temperature.
It follows from Eq.~\ref{Eq:ACFG} that in this case the ACF obtained as the result of an FCS experiment is:
\begin{equation}
G_\mathrm{SD} \left( \tau,w_0 \right) = \frac{1}{N} \left[ 1 +  \frac{\tau}{\tau_D} \right]^{-1} \left[ 1 +  \frac{1}{S^2} \frac{\tau}{\tau_D}  \right]^{-\frac{1}{2}},
\label{Eq:ACFSD}
\end{equation}
where $\tau_D = w_0^2 / \left( 4 D \right)$ is the characteristic decay time associated with the diffusion of the fluorophore through the observation volume.

\subsection{Fractional Brownian motion} \label{anomalous}

A frequently discussed anomalous diffusion process with a Gaussian propagator is fractional Brownian motion (FBM) \cite{Mandelbrot1968,Goychuk2009}. An FBM process has a mean-squared displacement $MSD(t) = 6At^{\alpha}$. It then follows from Eq.~\ref{Eq:ACFG} that:
\begin{equation}
G_\mathrm{FBM} \left( \tau,w_0 \right) = \frac{1} {N} \left[ 1 + \left( \frac{\tau}{\tau_D} \right) ^{\alpha} \right]^{-1} \left[ 1 + \frac{1}{S^2} \left(  \frac{\tau}{\tau_D} \right) ^{\alpha} \right] ^{-\frac{1}{2}},
\label{Eq:ACFAD}
\end{equation}
with a characteristic decay time $\tau_D = (w_0^2 / 4A)^{1/\alpha}$. This ACF is self-similar, meaning that if the lengthscale of the experiment (i.e. $w_0$, and thus $\tau_D$) is changed, the shape of the ACF does not change (as long as $S$ is kept constant). This property reflects the self-similarity of the FBM process, which does not have any characteristic timescale or lengthscale. 

It is important to recognize that, while often fit to diffusion data, Eq.~\ref{Eq:ACFAD} is strictly correct only for a diffusion process with a Gaussian propagator and a mean-squared displacement proportional to $t^{\alpha}$. However, it has been shown to be a good approximation for a large class of anomalous diffusion processes when studied with single-scale FCS \cite{Weiss2004}.

\section{Experimental} \label{methods}

\subsection{Reagents}

Fluorescent tracers (Alexa Fluor 488, Oregon Green 488, Streptavidin conjugated to Oregon Green, Streptavidin conjugated to Alexa 488, $40$ nm-diameter orange polystyrene FluoSpheres) were purchased from Invitrogen (now Life Technologies, Grand Island, NY). Dextran with a peak molecular weight of $276$ kDa and a low polydispersity index of $1.73$ was obtained from Sigma-Aldrich (St. Louis, MO), and biotechnology grade agarose from BioShop (Burlington, Canada).

\subsection{Sample preparation}

For calibration and experiments in aqueous solutions, the fluorescent tracers were simply dissolved in phosphate buffered saline (PBS, pH 7.4) at concentrations suitable for FCS (typically 1-10nM). In the case of the FluoSpheres, the solution was sonicated for $20$ min to reduce aggregation. 

Polymer solutions were prepared as previously described, by dissolving $276$ kDa dextrans at $200$ g/l in PBS \cite{Banks2005}. Agarose polymer gels were prepared by dissolving agarose in PBS at $90^{\circ}$C to produce $1.3\%$ solutions by weight. Both $500~{\mu}$l of the agarose solution and a $10~{\mu}$l concentrated tracer solution were then (separately) placed in a $60^{\circ}$C bath. After several minutes, both samples were mixed and vortexed, then returned to the $60^{\circ}$C bath and sonicated for $20$ min. Finally, $300~{\mu}$l of the solution was transferred to a pre-heated 96-well plate with coverslip bottom (Whatman, Clifton, NJ). Pre-heated micropipette tips were used during the whole procedure to prevent local cooling of the gel and large-scale spatial heterogeneities. The plate was covered with clear plastic wrap to prevent evaporation and placed in a $60^{\circ}$C oven for $10$ min. The sample was then allowed to cool to room temperature over the course of an hour before performing FCS measurements.

\subsection{VLS-FCS experiments}

The home-built FCS setup used for this work has been described previously \cite{Banks2005}. To perform VLS-FCS, a calibrated iris diaphragm was added to the parallel path of the $488$ nm excitation beam, midway between a parallel beam expander and the objective lens, in order to control the width of the beam. The size of the observation volume was increased by incremental reduction of the iris diameter concentric to the optical axis ($12$, $4$, $2.5$, $1.25$, $0.9$ and $0.5$ mm) and by selecting confocal pinholes with incrementally higher diameters ($50$, $75$, $100$, $150$, $200$, and $300~\mu$m). This resulted in observation volume radii evenly spaced on a logarithmic scale, $w_0 \simeq 0.3$ to $2.5~\mu$m. Using a variable focus lens as a beam expander (as done in other studies \cite{Wawrezinieck2005,Masuda2006}) would have automatically preserved the Gaussian profile of the excitation beam, but truncating the beam with an iris allowed reducing the size of the beam further and obtaining an increased range of observation lengthscales. Furthermore, the truncated excitation beam, having had to propagate for $\sim 20$ cm before reaching the objective lens, must have taken the form of a Laguerre-Gaussian beam, likely with a predominant $\text{TEM}_{00}$ Gaussian mode. The confocal observation volume is then expected to be indistinguishable from a three-dimensional Gaussian volume, as long as the confocal pinhole has a size adapted to that of the focal spot \cite{Hess2002}. With the pairings in beam size and confocal pinhole we used, we could detect no deviation from Gaussian for the observation volumes, as shown by the fact that the correlation data produced by organic fluorophores freely diffusing in PBS was very well fitted assuming free diffusion  
(with $S \sim 5 - 10$), even at the largest observation volumes. Although changing the optical index of the sample by adding crowding molecules might in principle cause optical distortions, previous works have shown that this was the case neither for the dextran solution \cite{Banks2005} nor for the agarose gel \cite{FatinRouge2004} we used.

An important aspect of the experiments was that the excitation power reaching the objective, $P$, was held constant (at $20~\mu$W for Streptavidin-Oregon Green, and $50~\mu$W for the more photostable FluoSpheres) using a continuously variable metallic neutral density filter. The probability for a fluorophore to photobleach during its passage through the observation volume is proportional to the excitation intensity, $\Phi$, and to the average residence time of the fluorophores in the observation volume, $\tau_D$. Since $\Phi = P/(\pi w_0^2)$ and $\tau_D \propto w_0^2$, a constant $P$ ensured that the probability to photobleach a fluorophore remained approximately constant when changing the size of the observation volume. The specific brightness of the diffusing fluorescent particles is expected to scale as $w_0^{-2}$ with a constant excitation power, meaning that it should be greatly reduced for large observation volumes, eventually limiting the accessible range of observation sizes. However, the fluorescence per molecule scaled as $w_0^{-1.8}$ for the organic dyes used in the study (e.g. varying from $\sim 30$ to $1.5$ kHz for Streptavidin-Oregon Green in PBS, when increasing observation volume size), and as $w_0^{-1.3}$ for the FluoSpheres (varying from $\sim 300$ to $25$ kHz in PBS). This effect, maybe due to fluorescence quenching (expected to be stronger for FluoSpheres, which are labeled with large numbers of individual fluorescent molecules, than for organic dyes) or saturation effects, allowed performing acceptable FCS measurements over the whole range of observation volumes considered.

For each observation volume size, $w_0$ was first determined by performing an FCS experiment with a dye with known diffusion coefficient (fluorescein, $D=425~\mu\text{m}^2/s$ at $25^{\circ}C$ (Ref.~\citenum{Culbertson2002}), or Alexa 488, $D=435~\mu\text{m}^2/s$ at $22.5^{\circ}C$ (Ref.~\citenum{Petrasek2008})). Then at least ten successive measurements (each $1$ to $3$ min in duration) were performed at the same position in each sample.

\subsection{Analysis of autocorrelation functions} \label{photo}

The ACFs were analyzed using an expression taking into account both fluorophore diffusion and photophysics: 
\begin{equation}
G \left( \tau, w_0 \right) = G_D  \left( \tau, w_0 \right) \times \prod_{i=1}^{i=n} \left( 1 + \frac{T_i}{1-T_i} e^{- \frac{\tau}{\tau_{T,i}}} \right).
\label{Eq:ACF}
\end{equation}
$G_D \left( \tau,w_0 \right)$ is the contribution of the fluorophore diffusion, assumed to be given by either  Eq.~\ref{Eq:ACFSD} or \ref{Eq:ACFAD}. The second term accounts for the existence of independent non-fluorescent states\cite{Widengren1995}. State $i$ has a relaxation time $\tau_{T,i}$ and a fractional occupation $T_{i}$. Fitting was performed with the analysis program Kaleidagraph (Synergy Software). It relies on a Levenberg-Marquardt algorithm to minimize the chi-square value, $\chi^2 = \sum \left[ G_\mathrm{exp}\left( \tau_i, w_0 \right) - G_\mathrm{th}\left( \tau_i, w_0 \right) \right]^2/\sigma_i^2$, where $G_\mathrm{exp}$ and  $G_\mathrm{th}$ are the experimental and theoretical values of the autocorrelation function, respectively. Unless otherwise specified, unweighted fits (\textit{i.e.}\ $\sigma_i=1$ for all data points) were performed.

\begin{figure}
\centering
  \includegraphics[width=7cm]{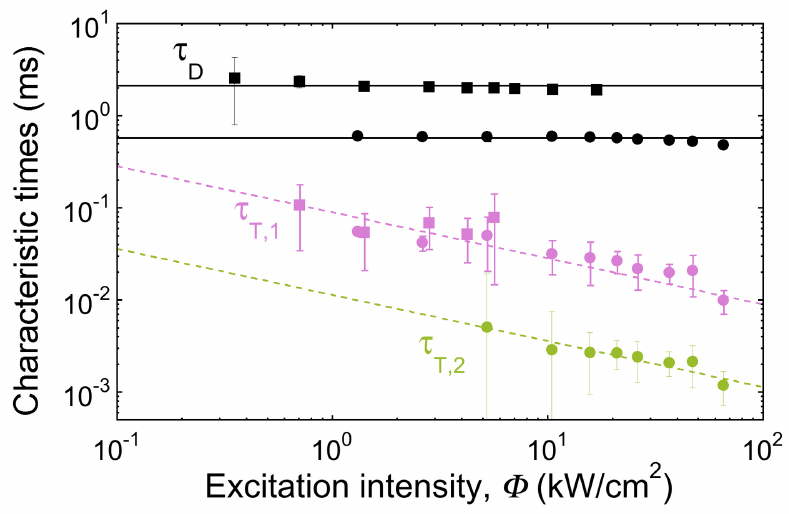}
  \caption{Characteristic times observed for Streptavidin-Oregon Green in buffer, as a function of the excitation intensity. The diffusion time ($\tau_D$, black symbols), the relaxation time for the triplet state ($\tau_{T,2}$, green symbols) and the relaxation time for an additional dark state ($\tau_{T,1}$, pink symbols) were obtained by fitting the ACFs with Eq.~\ref{Eq:ACF} (mean $\pm$ standard deviation). Data obtained for two different detection volume sizes are shown: $w_0 = 350$ nm (squares) and $w_0 = 675$ nm (circles). At low excitation intensity, short lag time noise in the ACF prevents determining the characteristic times of the photophysics terms reliably. Lines are fits to the data assuming either no dependence on the energy intensity (continuous lines) or a $\Phi^{-0.5}$ dependence (dashed lines).}
  \label{fgr:figure1}
\end{figure}

For the FluoSpheres, no photophysics term was observed, probably due to the large number of dye molecules in each bead. Thus, the correlation data was fitted with Eq.~\ref{Eq:ACF} using $n=0$. In the case of the organic dyes used for calibration, only one non-fluorescent state, a triplet state, was detected, and Eq.~\ref{Eq:ACF} was used with $n=1$. 
For fluorescently labeled Streptavidin an additional correlation term was observed, with a relaxation time $\tau_{T,1}$ that increased from $40~\mu$s to $\sim 1$~ms as the observation volume size was increased (the small amplitude of this term, $T_1 < 0.1$, made it difficult to obtain precise values of the relaxation time for large observation volumes). A dependence on the observation volume size might indicate a diffusion term. However, the characteristic time $\tau_{T,1}$ associated with this additional term was here also found to depend on the excitation intensity, $\Phi$, as expected for a photophysical process \cite{Widengren1995} (Fig.~\ref{fgr:figure1}). Since $\Phi$ decreased as the observation volume size was increased (as $P$ was kept constant), the apparent volume size dependence of $\tau_{T,1}$ is thus due to its underlying dependence on the excitation intensity. This evidences that a photophysical rather than a diffusion process is at the origin of the extra term observed in the correlation. This extra correlation was observed both when the protein was labeled with Alexa Fluor 488 and with Oregon Green, but not for the dyes alone. This further suggests that it is due to the properties of the protein (\textit{e.g.} proximity of tryptophan residues to the binding site of the dye resulting in quenching effects \cite{Buschmann2003,Mazouchi2013}) rather than to that of the dyes.  
The amplitude of the extra correlation was smallest when the protein was labeled with Oregon Green, therefore Streptavidin labeled with Oregon Green was used in all the experiments requiring a protein tracer. In that case, ACFs were fitted with Eq.~\ref{Eq:ACF} using $n=2$. 

Calibration measurements done with organic dyes were fit with Eq.~\ref{Eq:ACF} (assuming simple diffusion, i.e.\ using Eq.~\ref{Eq:ACFSD} for the diffusive part of the ACF), yielding values for both $w_0$ and~$S$. Unless otherwise specified, the correlation data obtained with the protein samples were then fit with Eq.~\ref{Eq:ACF} assuming FBM (i.e.\ using Eq.~\ref{Eq:ACFAD} for the diffusive part of the ACF), with  the value of $S$ fixed to that obtained in the calibration step, in order to retrieve $\alpha$ and $\tau_D$. To emphasize that the value of $\alpha$ obtained in this manner might differ from the true anomalous exponent of the diffusion if the propagator does not obey  Eq.~\ref{propagator}, we refer to it in the following as $\alpha_{ACF}$. In the case of measurements obtained for FluoSpheres, the FCS data often contained erratic correlations at lag times above $\tau_D$,  that may be attributed to residual aggregations and poor statistics. In order to obtain reliable fits, the data was weighted proportionally to the amplitude of the autocorrelation function (\textit{i.e.}\ $\sigma_i^2 = 1/G_\mathrm{exp}\left( \tau_i, w_0 \right)$), such that more weight was given to the more reliable data found at lag times below and around the characteristic diffusion time. This did not change the average values of $\alpha_{ACF}$ or $\tau_D$, but it resulted in a significant increase in the number of data sets with stable fits.

A second type of analysis was performed on the ACF obtained for the protein samples to test the nature of the diffusion, as suggested recently \cite{Horton2010,Hofling2011,Hofling2013}. After normalization of the experimental ACFs in order to bring the amplitude of the diffusive part to $1$, Eq.~\ref{Eq:ACFG} was used to perform a numerical inversion of the data at each lag time (using the software Mathematica), in order to extract the function $MSD \left( \tau \right)$ directly from the data. In the following, we refer to the function obtained in this manner as $\text{MSD}_{ACF} \left( \tau \right)$, to emphasize that it might differ from the actual mean-squared displacement if the diffusion propagator is not Gaussian.

\subsection{Analysis of diffusion laws}

For each sample, the diffusion law, that is the comparison of the value of the characteristic diffusion time extracted from the ACF, $\tau_D$, with the radius of the observation volume calculated from the dye calibration experiment, $w_0$, was examined to test the scaling expected for anomalous diffusion:
\begin{equation}
w_0^2 = 4 A \tau_D ^ {\alpha}.
\label{Eq:Law1}
\end{equation}
The value of $\alpha$ obtained from fitting Eq.~\ref{Eq:Law1} to the experimental diffusion law is referred to in the following as $\alpha_{DL}$ to emphasize that it was obtained with the assumption of a power law relationship between $w_0$ and $\tau_D$.
For comparison with previous studies, the same data was also fitted to a non-Brownian linear diffusion law \cite{Wawrezinieck2005, Wawrezinieck2004} : 
\begin{equation}
\tau_D = \frac{1}{4D} w_0^2 + \tau_0.
\label{Eq:Law2}
\end{equation}
This allowed checking for the presence of a time intercept, $\tau_0$, shown to be linked to the existence of confinement \cite{Wawrezinieck2005}.

\section{Results} \label{results}

\subsection{Protein diffusion in a crowded dextran solution}

\subsubsection{Buffer. } 
\begin{figure}[!t]
\centering
  \includegraphics[width=8cm]{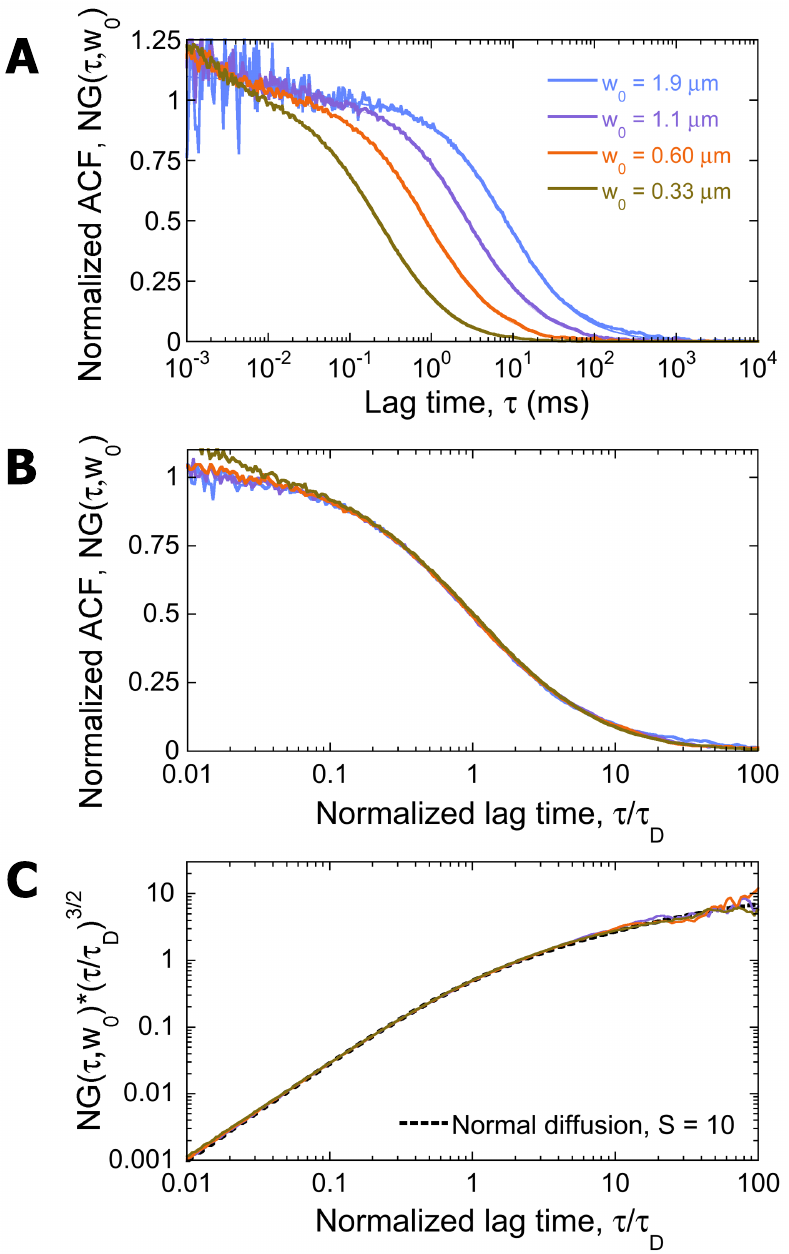}
  \caption{(A) Average ACFs obtained for Oregon Green-labelled Streptavidin diffusing in PBS for different sizes of the observation volume (thick lines). Fits assuming simple diffusion (Eqs.~\ref{Eq:ACF} and~\ref{Eq:ACFSD}, i.e.\ with $\alpha_{ACF}$ fixed to $1$, and with $S=10$ and $n = 2$) are shown (thin lines, almost indistinguishable from the data). Both the ACFs and their fits have been normalized by $1/N$ to make the amplitude of the diffusion component equal to $1$.  (B) Same ACFs as in (A), after normalization of the lag time by the characteristic diffusion time, $\tau_D$. (C) Three of the ACFs shown in (B), after rescaling by $\left( \tau/ \tau_D \right) ^{3/2}$ (the ACF recorded for $w_0 = 1.9~\mu \text{m}$ was omitted because of noise at long lag time). The data is compared to the curve expected for free diffusion in the absence of photophysics (Eq.~\ref{Eq:ACF} with $\alpha_{ACF}=1$, $S = 10$ and $n=0$, dashed black line).}
  \label{fgr:figure2}
\end{figure}

We first verified that VLS-FCS data acquired for the diffusion of a tracer protein in buffer was thoroughly consistent with simple diffusion. Streptavidin conjugated to Oregon Green was diluted in PBS at a nanomolar concentration, and FCS experiments were performed on this solution for a series of observation volumes. ACFs obtained for several different values of $w_0$ are shown in Fig.~\ref{fgr:figure2}A (each curve is the average of $5$ ACFs obtained in the same conditions for the same sample). The correlation data was fitted assuming the presence of two photophysical processes (Eq.~\ref{Eq:ACF} with $i = 2$), as explained in section \ref{photo}, and using the value $S = 10$ obtained from calibration measurements. As shown in Fig.~\ref{fgr:figure2}A, the ACFs are very well fitted using a simple diffusion model, \textit{i.e.}\ fixing $\alpha_{ACF} = 1$. ACFs after amplitude normalization (using the value of $N$ obtained from the fit) and lag time normalization (using the characteristic diffusion time, $\tau_D$, obtained from the fit) are shown in Fig.~\ref{fgr:figure2}B. The normalized ACFs obtained at different observation volume sizes overlap perfectly for almost four decades centred around $\tau_D$, except at very short lag times, where photophysics effects start playing a role for ACFs obtained at small detection volumes, and at very long lag times, where noise due to insufficient statistics starts playing a role for ACFs obtained at large detection volumes. This self-similarity illustrates the unchanging properties of diffusion as the lengthscale of the experiment is varied, as expected for simple diffusion. For better visualization of the asymptotic behaviour of the ACFs, they are shown in Fig.~\ref{fgr:figure2}C after rescaling by $\left( \tau/ \tau_D \right) ^{3/2}$. Although the correlation data becomes too noisy at large lag times to observe real asymptotic behaviour, the ACFs collapse nicely over four orders of magnitude in correlation, and seem to approach a plateau at large lag times as expected for simple diffusion.

\begin{figure}[!t]
\centering
  \includegraphics[width=8cm]{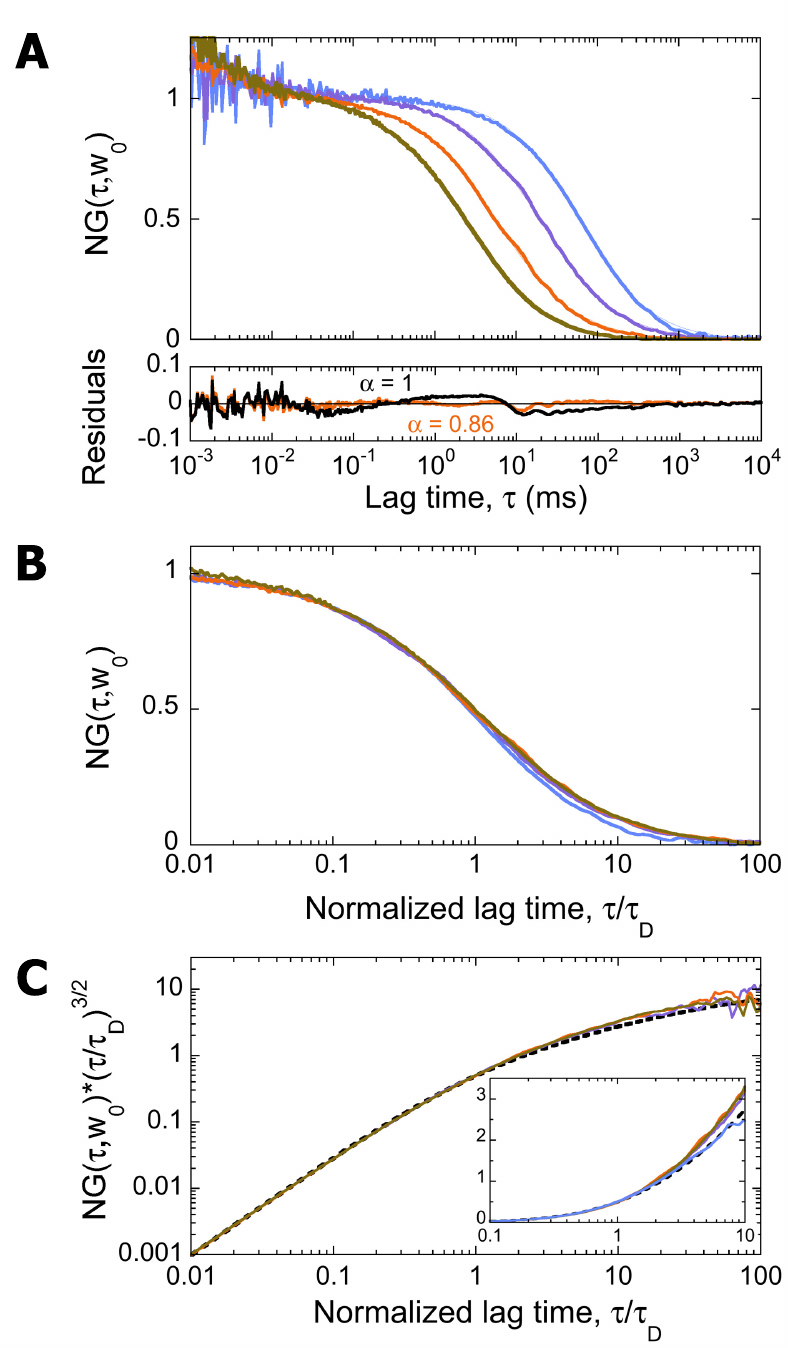}
  \caption{(A) Average ACFs obtained for Streptavidin-Oregon Green diffusing in PBS crowded with $276k$ Da dextran at $200$ g/l, for different sizes of the observation volume (thick lines, same colour code as in Fig.~\ref{fgr:figure2}). Fits assuming anomalous diffusion (Eqs.~\ref{Eq:ACF} and~\ref{Eq:ACFAD} with $S=10$ and $n = 2$) are shown (thin lines). Both the ACFs and their fits have been normalized by $1/N$. The residuals of the fits assuming either simple or anomalous diffusion are shown for the ACF acquired at $w_0 = 0.60~\mu \text{m}$. (B) Same ACFs as in (A), after normalization of the lag time by $\tau_D$. (C) Same ACFs as in (B), after rescaling by $\left( \tau/ \tau_D \right) ^{3/2}$. The inset shows the same curves on a linear-log scale for lag times around $\tau_D$ (the ACF recorded for $w_0 = 1.9~\mu \text{m}$ was omitted in the main panel because of noise at large lag time, but is shown in the inset). The data is compared to the curve expected for free diffusion in the absence of photophysics (Eq.~\ref{Eq:ACF} with $\alpha_{ACF}=1$, $S = 10$ and $n=0$, dashed black line).}
  \label{fgr:figure3}
\end{figure} 

\subsubsection{Crowded polymer solution. } The results obtained for Streptavidin-Oregon Green diffusing in a concentrated (200 g/l) solution of 276 kDa dextran dissolved in PBS (Fig.~\ref{fgr:figure3}) were in contrast to those observed for the protein in PBS. The ACFs obtained for crowded dextran solutions were consistently better fitted when assuming anomalous diffusion, \textit{i.e.}\ letting $\alpha$ vary in Eq.~\ref{Eq:ACF} (Fig.~\ref{fgr:figure3}A). This had been reported previously for a single FCS lengthscale \cite{Banks2005,Sanabria2007,Szymanski2009}. More surprising, however, is that the overall shape of the ACFs changes slightly with observation lengthscale (Fig.~\ref{fgr:figure3}B), when one would expect self-similarity for the simplest anomalous diffusion models, such as the FBM model presented in section \ref{anomalous}. As the lengthscale is reduced, the ACFs broadens, indicating that the diffusion becomes more and more anomalous. The same conclusions can be reached when looking at the data rescaled with $\left( \tau/ \tau_D \right) ^{3/2}$ (Fig.~\ref{fgr:figure3}C). The asymptotic behaviour of the ACFs deviates from that expected for simple diffusion, in a slightly scale-dependent manner.

\begin{figure}[!t]
\centering
  \includegraphics[width=6.5cm]{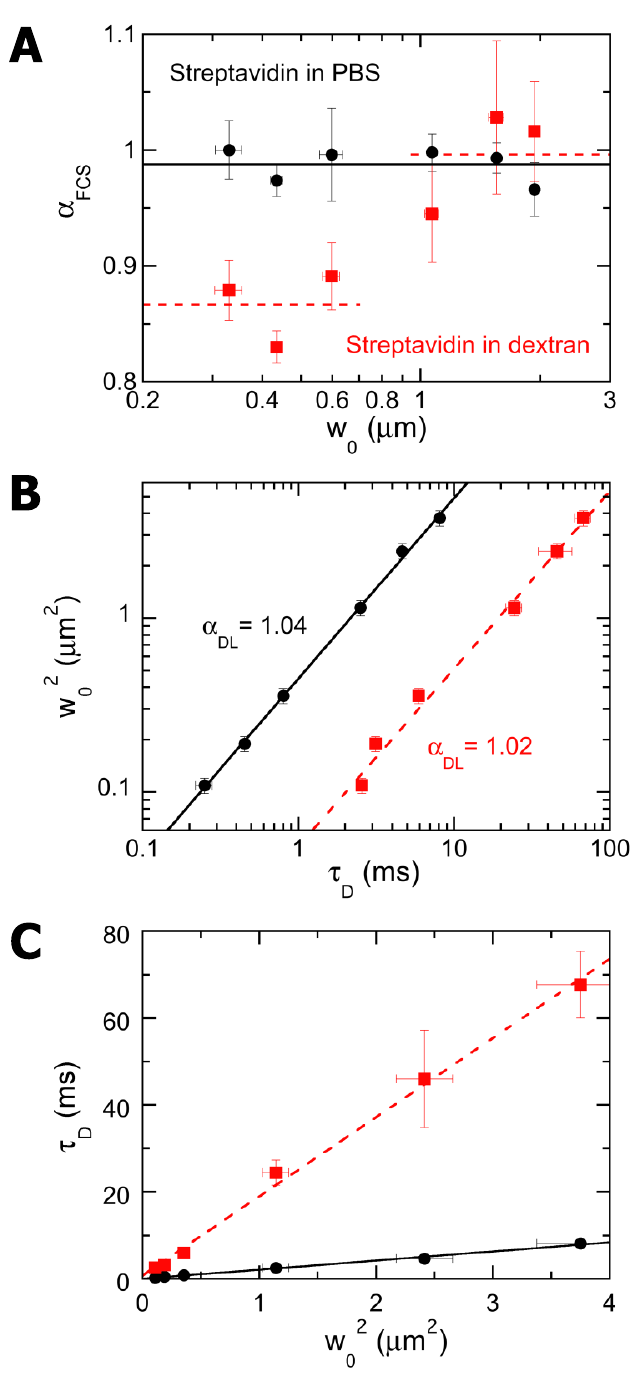}
  \caption{ (A) Values of $\alpha_{ACF}$ (mean $\pm$ standard deviation) determined for Streptavidin-Oregon Green in buffer (black circles) and in buffer crowded with 200 g/l 276 kDa dextran (red squares) by fitting the ACFs with Eqs.~\ref{Eq:ACF} and~\ref{Eq:ACFAD}. (B) Diffusion laws obtained for the same samples (filled symbols), plotted on a log scale, and fitted with Eq.~\ref{Eq:Law1} to extract the value of $\alpha_{DL}$ (lines). (C) Same as in (B), but plotted on a linear scale. Lines are fits of Eq.~\ref{Eq:Law2} to the data, showing that the intercept at $w_0 = 0$ is $\tau_0 \simeq 0$, and returning $D = 119~\mu m^2/s$ and $D = 14~\mu m^2/s$ for the diffusion coefficient of the protein in the absence and in the presence of dextran crowders, respectively. }
  \label{fgr:figure4}
\end{figure}

\begin{figure}[!t]
\centering
  \includegraphics[width=7cm]{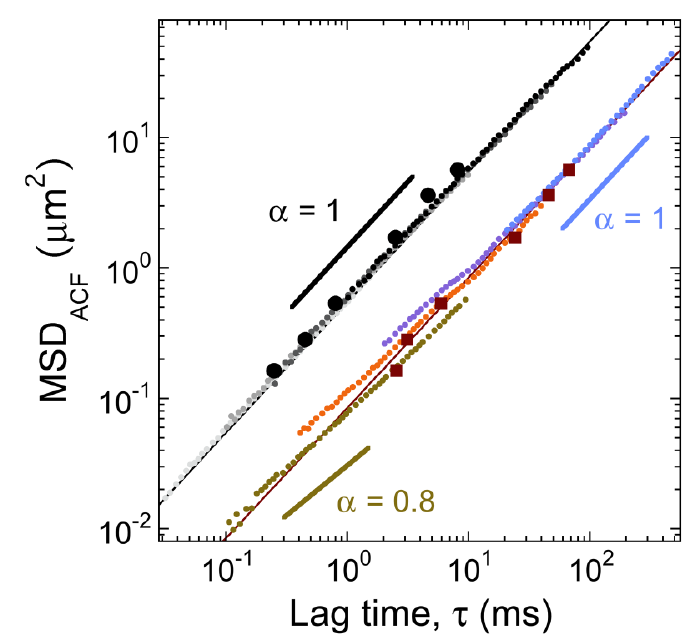}
  \caption{Mean-squared displacement of diffusing Streptavidin-Oregon Green extracted from the ACFs shown in Figs.~\ref{fgr:figure2} and ~\ref{fgr:figure3}, using the inversion procedure described in section~\ref{inversion}. The data obtained in the absence (small grey and black symbols) and in the presence (small coloured symbols) of dextran crowding are compared. Solid lines indicate the slopes expected for different values of the anomalous exponent. Fits to the data assuming simple diffusion (i.e.\ a Gaussian propagator and $\alpha=1$) are indicated with thin lines and return values of the diffusion coefficient of Streptavidin of $D = 90~\mu \text{m}^2/s$ in the absence of dextran crowders and $D = 14~\mu \text{m}^2/s$ in its presence. For comparison, large symbols show the values of $3 w_0^2 / 2$ as a function of $\tau_D$, \textit{i.e.}\ the diffusion law data from Fig.~\ref{fgr:figure4}B (black circles: in the absence of dextran, red squares: in the presence of dextran). }
  \label{fgr:figure5}
\end{figure}

The average values of the anomalous exponents (obtained for several repeats of the same experiment at each lengthscale for the diffusion of streptavidin) in both PBS and dextran are compared in Fig.~\ref{fgr:figure4}A. As discussed above, $\alpha_{ACF} \simeq ~1$ at all lengthscales for diffusion in PBS, while $\alpha_{ACF}$ varies from $\simeq 0.85$ for small observation volumes (the error is rather large since the second photophysics term has a characteristic time close to that of the diffusion) up to $\simeq 1$ for the larger observation volumes. The diffusion law, \textit{i.e.}\ the relationship between lengthscale ($w_0$) and timescale ($\tau_D$), is shown in Figs.~\ref{fgr:figure4}B and C. We first tested whether their might be a deviation from a linear scaling with time, as predicted by most anomalous diffusion models and as suggested by the ACFs recorded in the presence of polymer crowding. This was done by fitting the diffusion law with Eq.~\ref{Eq:Law1} (Fig.~\ref{fgr:figure4}B). For diffusion in both PBS and in the polymer solution, the result was too close from linearity to be deemed anomalous ($\alpha_{DL} = 1.04 \pm 0.04$ in PBS and $\alpha_{DL} = 1.01 \pm 0.04$ in the crowded polymer solution). This means that Streptavidin has a lengthscale-independent diffusion coefficient (at the studied lengthscales) both in PBS and in the dextran solution. In the case of the polymer solution, this result is in apparent contradiction with the $\alpha_{ACF} < 1$ values measured at small lengthscales. Next, assuming that there was a linear relationship between $\tau_D$ and $w_0^2$, as done for example in Ref.~\citenum{Wawrezinieck2005}, we fitted the experimental diffusion laws with Eq.~\ref{Eq:Law2} (Fig.~\ref{fgr:figure4}C). Within error, there was no intercept ($\tau_0 \simeq 0$) for either system.

To further explore the discrepancy in anomalous exponent uncovered between individual ACFs and the diffusion law, we used an ACF inversion method originally proposed by Shusterman \textit{et al.} \cite{Shusterman2004,Shusterman2008}, and later put forward in the context of anomalous diffusion studies \cite{Horton2010,Hofling2011,Hofling2013}. This method allows extracting a function directly from the ACF, $\text{MSD}_{ACF} \left( \tau \right)$, which corresponds to the mean-squared displacement of the particles as long as their diffusion is governed by a Gaussian propagator (see section \ref{inversion}). It does require a precise knowledge of $w_0$, which in our case was obtained at each lengthscale from a separate calibration experiment using a dye with known diffusion coefficient. The result of this inversion is shown in Fig.~\ref{fgr:figure5}. Each ACF provides reliable data for about two decades in time around the characteristic diffusion time (limited by the dye photophysics at small lag times, and by statistical noise at large lag times). When stitched together, the data extracted from the ACFs obtained at different observation volumes allowed exploring the MSD for more than four decades in time. If the Gaussian propagator assumption was correct, then the $\text{MSD}_{ACF} \left( \tau \right)$ extracted at different scales should overlap. This is clearly the case for the diffusion of the protein in PBS buffer, for which all the $\text{MSD}_{ACF} \left( \tau \right)$ also have a slope of $1$, consistent with simple diffusion. However, in the case of the diffusion of the protein in the crowded dextran solution, the different $\text{MSD}_{ACF} \left( \tau \right)$ do not overlap. Each individual $\text{MSD}_{ACF} \left( \tau \right)$ has a different slope (decreasing from $1$ to $\approx 0.8$ as the scale is reduced, in agreement with the values of $\alpha_{ACF}$ obtained from the correlation data, see Fig.~\ref{fgr:figure4}A). Yet overall, they seem to align along a line with slope $1$ (in agreement with the diffusion law showing a linear dependence between $w_0^2$ and $\tau_D$, see Fig.~\ref{fgr:figure4}B). Interestingly, for those $\text{MSD}_{ACF} \left( \tau \right)$ encompassing $\tau \simeq 10~\text{ms}$, a clear kink was observed at that timescale, showing a rather clear delimitation between simple diffusion (above $\simeq 10~\text{ms}$ or $\simeq 1~\mu$m) and anomalous diffusion (below $\simeq 10~\text{ms})$.

\subsection{Bead diffusion in agarose gel}

\begin{figure}[!t]
\centering
  \includegraphics[width=8cm]{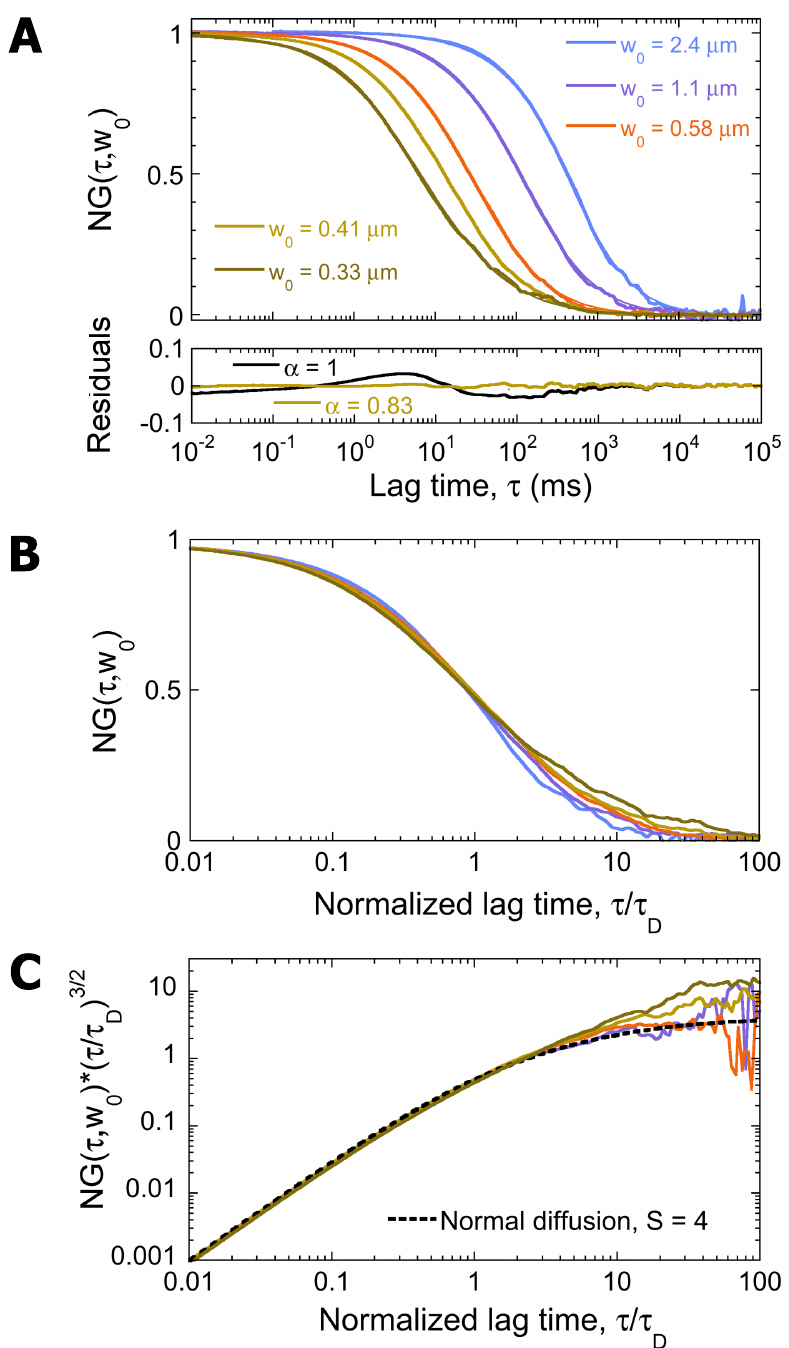}
  \caption{ (A) Average ACFs acquired for 40 nm fluorescent beads diffusing in a $1.3\%$ agarose gel, for different sizes of the observation volume (thick lines). Fits assuming anomalous diffusion (Eqs.~\ref{Eq:ACF} with $n = 0$ and Eq.~\ref{Eq:ACFAD}, $S = 4$ and $\alpha$ left to vary) are shown (thin lines). Both the ACFs and their fits have been normalized by $1/N$. The residuals of the fits with simple and anomalous diffusion are shown for the ACFa acquired at $w_0 = 0.41~\mu\text{m}$ in the lower panel. (B) Same ACFs as in (A), after normalization of the lag time by the characteristic diffusion time, $\tau_D$. (C) Four of the ACFs shown in (B), after rescaling by $\left( \tau/ \tau_D \right) ^{3/2}$ (the ACF recorded for $w_0 = 2.4~\mu \text{m}$ was omitted because of excessive noise at long lag times). The data is compared to the curve expected for free diffusion in the absence of photophysics ($\alpha_{ACF}=1$, $S = 4$ and $n=0$, dashed black line).}
  \label{fgr:figure6}
\end{figure}

\begin{figure}[!t]
\centering
  \includegraphics[width=6.5cm]{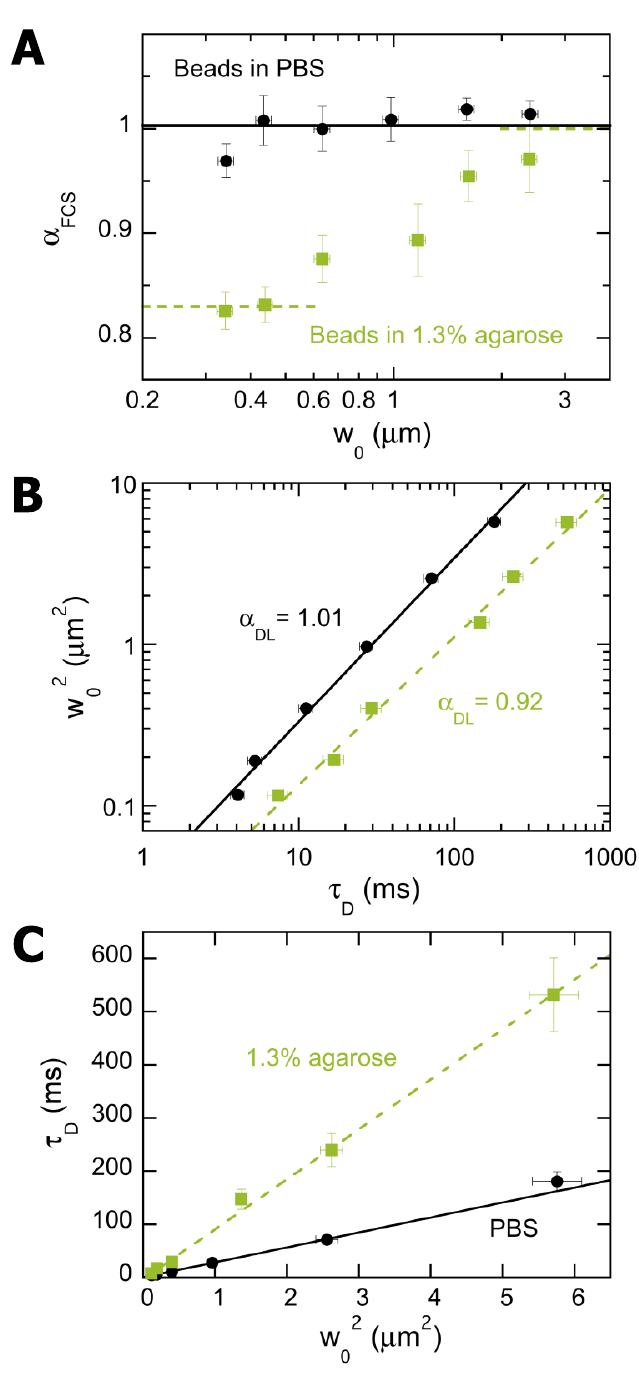}
  \caption{
 (A) Values of $\alpha_{ACF}$ determined for fluorescent beads in buffer (black circles) and in a $1.3 \%$ agarose gel (green squares) by fitting the ACFs with Eqs.~\ref{Eq:ACF} and~\ref{Eq:ACFAD}. (B) Diffusion laws obtained for these same samples. Lines are fit to the data with Eq.~\ref{Eq:Law1} to extract the values of $\alpha_{DL}$. (C) Same as in (B), but plotted on a linear scale, and fitted with Eq.~\ref{Eq:Law2}, showing that the intercept at $w_0 = 0$ is $\tau_0 \simeq 0$, and returning $D = 8.9~\mu \text{m}^2/s$ and $D = 2.7~\mu \text{m}^2/s$ for the diffusion coefficient of the beads in buffer and in the presence of agarose, respectively.}
  \label{fgr:figure7}
\end{figure}

Because of the ambiguous nature of the diffusion in the crowded polymer solution, we next looked at the diffusion of polystyrene beads in agarose gels, a system that had been consistently reported to show anomalous diffusion using both single particle tracking \cite{Valentine2001} and FCS \cite{Lead2003,FatinRouge2004,FatinRouge2006,Labille2007}. Again using VLS-FCS, we compared the diffusion of $40~\text{nm}$ beads in PBS with their diffusion in $1.3\%$ agarose gels. Average ACFs obtained for a representative FCS experiment with the beads in agarose gels are shown in Fig.~\ref{fgr:figure6}. The ACFs required fitting with an anomalous diffusion model (Fig.~\ref{fgr:figure6}A). Interestingly, as in the case of the polymer solution, the degree of anomaly depended on the lengthscale, as can be seen in the progressive broadening of the ACF and change in the asymptotic behaviour as $w_0$ becomes smaller (Fig.~\ref{fgr:figure6}B and C). 

This behaviour is consistent with the scale dependence of $\alpha_{ACF}$ obtained directly from the fit of the ACF, as shown in Fig.~\ref{fgr:figure7}A. Whereas  $\alpha_{ACF} \simeq 1$ at all lengthscales for beads in solution, it decreases in the case of the agarose gels from $\simeq 1$ at the largest (micron-size) observation volumes to $\simeq 0.85$ for the smaller (submicron-size) observation volumes achieved here. Looking at the experimental diffusion laws  (Fig.~\ref{fgr:figure7}B), a slight difference can this time be detected between the diffusion in PBS, for which $\alpha_{DL} = 1.01 \pm 0.3$ and the diffusion in the agarose gel, for which $\alpha_{DL} = 0.92 \pm 0.3$. A zero intercept was confirmed in both conditions (buffer and agarose gel) for these diffusion laws (Fig.~\ref{fgr:figure7}C).

\begin{figure}
\centering
  \includegraphics[width=7cm]{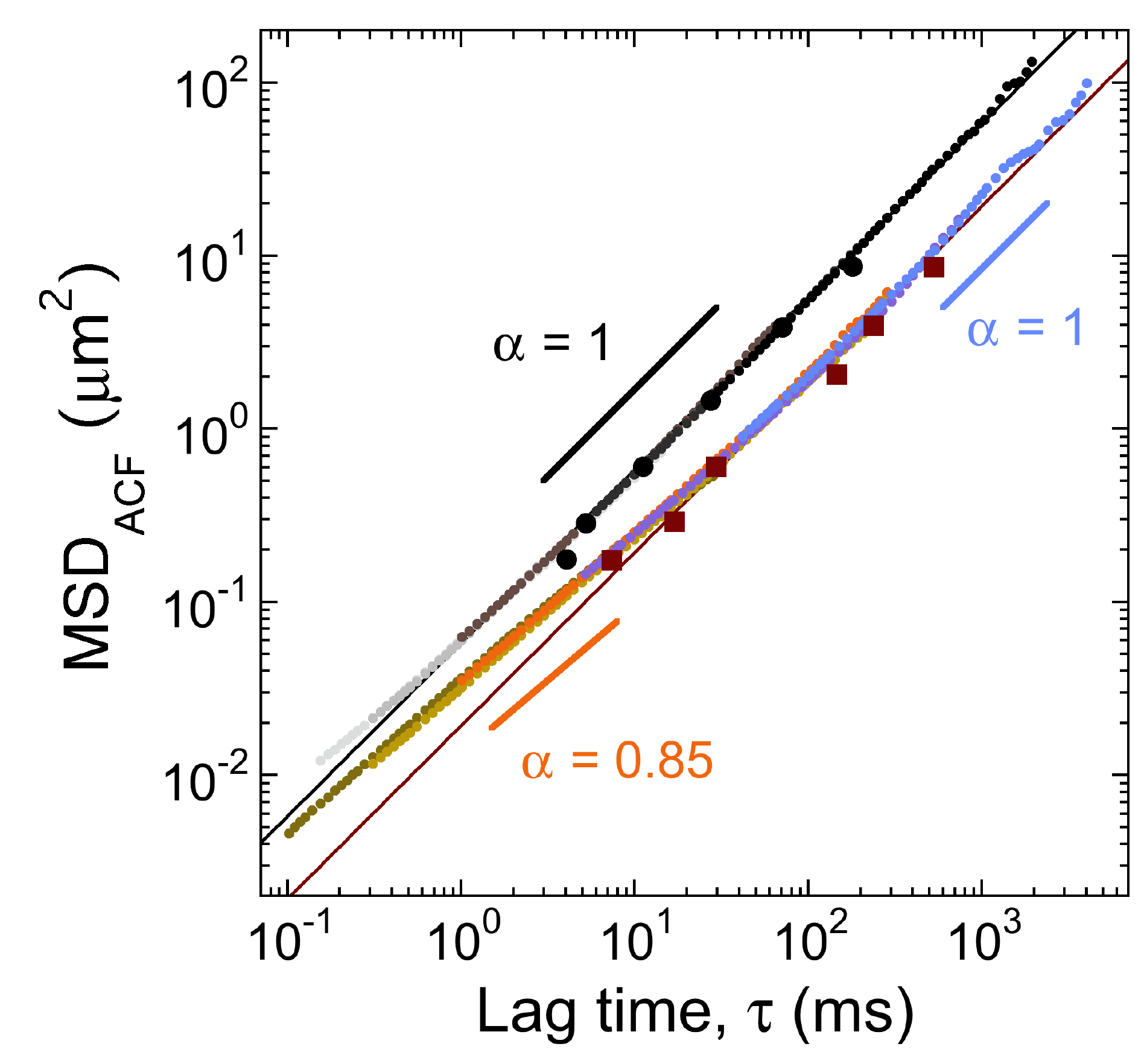}
  \caption{Mean-squared displacement of diffusing fluorescent beads in an aqueous buffer (black and grey symbols) and in a $1.3\%$ agarose gel, extracted from ACFs using the inversion procedure described in section~\ref{inversion} (coloured symbols). Solid lines indicate the slopes expected for different values of the anomalous exponent, $\alpha_{ACF}$. Fits of the data assuming simple diffusion (i.e.\ a Gaussian propagator and $\alpha=1$) return a diffusion coefficient for the beads $D = 9.6~\mu \textrm{m}^2/\textrm{s}$ in solution and $D = 3.2~\mu \textrm{m}^2/\textrm{s}$ in the agarose gel (thin lines). For comparison, large symbols show the values of $\frac{3}{2}w_0^2$ as a function of $\tau_D$, \textit{i.e.}\ the diffusion law data from Fig.~\ref{fgr:figure7}B (black circles: in buffer, red squares: in agarose).}
  \label{fgr:figure8}
\end{figure}

Remarkably, and in stark contrast with what was observed in the polymer solutions, when extracting $\text{MSD}_{ACF} \left( \tau \right)$ from the ACFs, it was found that the functions obtained at different scales overlapped almost perfectly, producing an MSD over 5 decades in time that curves from a slope $\alpha_{ACF} \approx 0.85$ at the smallest scales to $\alpha_{ACF} \approx 1$ at the largest, with a cross-over around $\tau \simeq 100~\textrm{ms}$ (Fig.~\ref{fgr:figure8}). This allows better understanding the results obtained by fitting Eq.~\ref{Eq:ACFAD} to individual ACFs (Fig.~\ref{fgr:figure7}). The diffusion law (Fig.~\ref{fgr:figure7}B) is fully consistent with $\text{MSD}_{ACF} \left( \tau \right)$ (Fig.~\ref{fgr:figure8}), and in fact exhibits the same cross-over that is clearly visible in $\text{MSD}_{ACF} \left( \tau \right)$. The value $\alpha_{\text{DL}} = 0.92$ obtained from the diffusion law is thus an average between the values of $\alpha$ before ($\alpha \simeq 0.85$) and after ($\alpha \simeq 1$) the cross-over. In full agreement with the cross-over interpretation, the values of $\alpha_{\text{FCS}}$ extracted from the ACFs go from $\alpha_{\text{FCS}} \simeq 0.85$ to  $\alpha_{\text{FCS}} \simeq 1$ (Fig.~\ref{fgr:figure7}A).

\section{Discussion}

\subsection{On the importance of characterizing complex diffusion processes over a range of lengthscales}

The results of many FCS and FRAP studies are interpreted in terms of a diffusion constant depending on lengthscale \cite{Banks2005,FatinRouge2004,Weiss2004,Weiss2003,Brown1999,Wachsmuth2000,Feder1996,Guigas2007}. However, in their simplest form these techniques have a fixed lengthscale, and thus do not provide a stringent test for anomalous diffusion. In that context, our study demonstrates the interest of performing variable-lengthscale experiments in order to refine the diagnostics of anomalous diffusion. It highlights the fact that variable-lengthscale FCS experiments can bring information on both the propagator and the MSD of the studied diffusion process. The MSD can be obtained from the diffusion law generated by repeating the FCS experiments at different scales \cite{Wawrezinieck2005}. The ACF, which directly depends on the propagator, reflects both its spatial form and its temporal evolution \cite{Lubelski2009,Hofling2013}. Although the propagator cannot be extracted directly from an experimental ACF, an assumption can be made about its Gaussianity, and, as discussed below, VLS-FCS data can be used to test this assumption.

\begin{figure}[!t]
\centering
  \includegraphics[width=8cm]{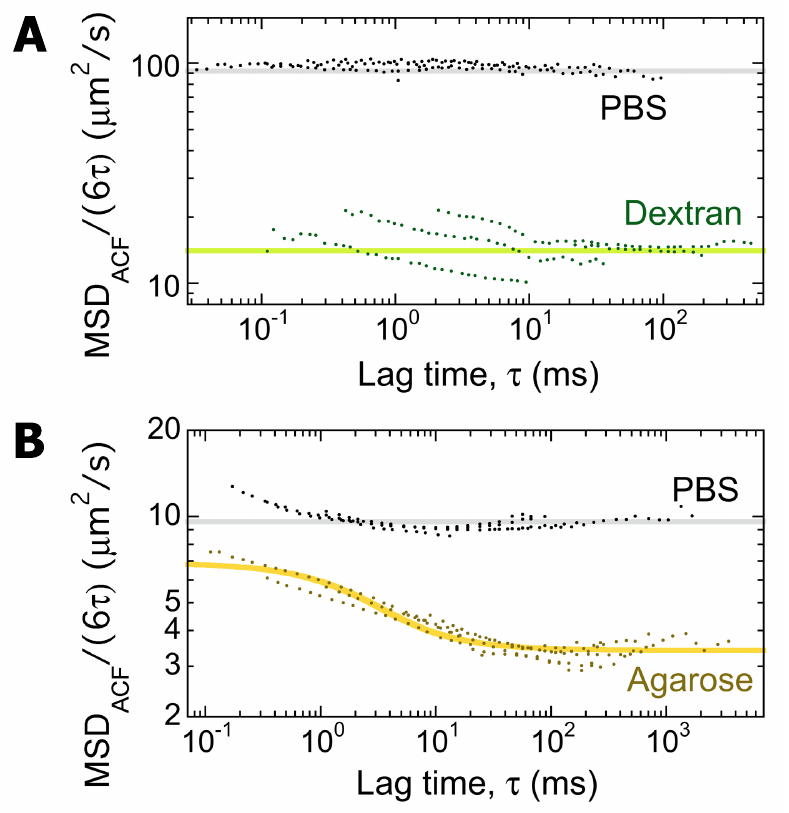}
  \caption{Mean-squared displacement divided by time for (A) Streptavidin-Oregon Green in PBS (black symbols) or in concentrated Dextran solution (green symbols) and (B) FluoroSpheres in PBS (black symbols) or in agarose gel (brown symbols). Horizontal lines indicate the expected form for simple diffusion with $D = 90~\mu \text{m}^2/s$ (dark grey line), $D = 14~\mu \text{m}^2/s$ (green line), or $D = 9.6~\mu \text{m}^2/s$ (light grey line). The orange line is a fit to a simplified model where one assumes diffusion through a series of semi-permeable cages of well defined size, giving: $MSD(\tau)/(6 \tau) =  L^2/(12 \tau) (1-exp(-\tau/\tau_{\text{eq}})) +D_\text{M}$\cite{Destainville2008}. This fit returns the long-time diffusion coefficient, $D_M = 3.4~\mu \text{m}^2/s$, the equilibration time,  $\tau_{\text{eq}} = 1.5~\text{ms}$ and the size of the cages, $L = 0.25~\mu \text{m}$. }
  \label{fgr:figure9}
\end{figure}

Several strategies have been employed for varying the size of the FCS observation volume, notably decreasing the size of the incoming excitation beam with a variable beam expander \cite{Masuda2005} or (as done in this study, allowing to cover almost an order of magnitude in lengthscales) with a variable diameter iris \cite{Wawrezinieck2005}. For fluorophores restricted to two-dimensional diffusion, the observation lengthscale can also be changed by placing the diffusion plane slightly out of focus \cite{Humpolivckova2006}, or by using the information captured in time-lapsed images \cite{Bag2012,Di2013,Di2014}. Recently, the possibility to extend the range of FCS measurements to sub-diffraction observation volumes has been demonstrated by a number of groups, using stimulated emission depletion\cite{Kastrup2005,Eggeling2008,Mueller2012,King2014}, zero-mode waveguides \cite{Samiee2006} or near-field optical probes \cite{Vobornik2008,Manzo2011}. Beyond the range of accessible observation volume sizes, our study emphasizes the importance of data quality, since it determines the range of lengthscales around $w_0$ for which the numerical inversion procedure to recover the MSD can be done. In favourable cases, we see that this procedure allows extending the accessible lengthscale range from $1$ to more than $4$ decades (see Fig.~\ref{fgr:figure8}, and Fig.~\ref{fgr:figure9} where the inverted FCS data is shown as a log-log plot of $MSD(\tau)/\tau$).

One important point highlighted by this study is that, although often used as a diagnostic for anomalous diffusion, the anomalous exponent extracted from ACFs, $\alpha_{ACF}$, is not sufficient to characterize complex diffusion processes. The complete shape of the ACF, which we argue is best taken in when inverted to generate an MSD plot (as in Figs. \ref{fgr:figure5} and \ref{fgr:figure8}), is much more informative, in particular because it can identify the presence of a cross-over. This inversion procedure is valid only in the case of a diffusion process with a Gaussian propagator. In media with a particular range of domain sizes, such as the crowded media studied here, it can be used safely for observation volumes that are either much larger or much smaller than the size of the domains, but its use must be critically assessed for intermediate observation volume sizes. Indeed, an interesting aspect of VLS-FCS is that the validity of the inversion procedure can be verified after the fact, by checking whether the inverted ACFs recorded at different lengthscales overlap (they should in the case of a Gaussian process). It thus provides a way to test the Gaussianity of the propagator at specific lengthscales. We note that the validity of making Gaussian approximations (such as the inversion procedure we used here) for diffusion in cross-over regions has recently been discussed in the context of single-particle tracking of diffusing colloidal particles \cite{Thorneywork2016}. We also note that whether diffusion in complex media is characterized by Gaussian propagators is a current topic of interest \cite{Kwon2014,Phillies2015,Ghosh2016}. However most of the data available so far has been obtained from single-particle experiments, and therefore is restricted to slow diffusion processes \cite{Wang2009, Wang2012,Xue2016}.

Obtaining information on both the propagator and the MSD is important to discriminate between different models of anomalous diffusion, which are distinguished by their different predictions for these two functions. Gaussian models of anomalous diffusion, such as the FBM model, have a Gaussian propagator combined with an MSD that is sublinear in time \cite{Sebastian1995,Ernst2012}. Others, like the continuous time random walk (CTRW) model\cite{Klafter1980,Lubelski2009} or the obstructed diffusion model \cite{Saxton1994,Havlin1987,Hofling2006,Franosch2011,Spanner2013} have neither a Gaussian propagator nor a linear MSD. Recently, a ``diffusing diffusivity'' model was proposed to explain diffusion processes with a linear MSD and a non-Gaussian propagator \cite{Chubynsky2014}.

\subsection{The caged diffusion of beads in agarose gels}

In the case of the diffusion of beads in agarose gel (used here as a positive control for anomalous diffusion \cite{Valentine2001,FatinRouge2004,Labille2007}), both the shapes of the individual ACFs and the diffusion law obtained by VLS-FCS clearly indicate a deviation from simple diffusion. The anomalous exponent recovered from the diffusion law, $\alpha_{DL} = 0.92$, falls within the range of those obtained from the individual ACFs, $\alpha_{ACF} \simeq 0.8 - 1$ (Fig.~\ref{fgr:figure7}A). Moreover, the extraction of the MSD from each individual ACF under the assumption of Gaussian diffusion yields a consistent MSD over more than five orders of magnitude in time (Fig.~\ref{fgr:figure8}). Our data thus confirms that diffusion of beads in agarose gels is anomalous, and further suggests that in this particular system the anomalous diffusion process is characterized by a Gaussian propagator. Not surprisingly, this last result differs from that reported for much larger tracers ($500~\text{nm}$ beads) in less dense agarose gels ($0.36\%$ by weight) \cite{Valentine2001}. In that case, the tracer particles became trapped at long observation times, indicating a very different diffusion process, and the average propagator measured by single particle tracking had an exponential (non-Gaussian) shape. Interestingly, a varying $\alpha_{ACF}$ (Fig.~\ref{fgr:figure7}A) points to the existence of a cross-over between different diffusion regimes, and therefore of a characteristic lengthscale in the agarose gel. A cross-over is indeed clearly visible in the continuous MSD shown in Fig.~\ref{fgr:figure8}, where a distinct change in slope occurs below a timescale of $\simeq 100$ ms, i.e. a lengthscale of $\simeq 1~\mu$m. The existence of a cross-over explains why a clear deviation from normal diffusion is observed when looking at the diffusion law on a log-log scale (Fig.~\ref{fgr:figure7}B), but not when looking at the same diffusion law on a linear scale where the data collected at larger scales takes more importance (Fig.~\ref{fgr:figure7}C).

In agarose gels, agarose fibres form a network, cross-linked by hydrogen bonds, defining pores with a distribution of sizes, through which particles that are small enough can diffuse. For agarose gels with concentrations around 1.5 wt\%, different values have been reported for the average pore diameter (which depends on temperature, cooling rate, pH, and buffer ionic strength), ranging from $\sim 80$ nm to $\sim 600$ nm \cite{FatinRouge2004,Narayanan2006,Pluen1999,Maaloum1998}. Our observations are consistent with momentary trapping of the beads in the pores formed by the agarose fibres, in which case one expects a cross-over from simple to anomalous diffusion to occur, at or slightly above the characteristic pore size. Different analytical models have been proposed to describe the cross-over in the MSD for arrays of cages \cite{Powles1992,Saxton2007,Destainville2008}. For simplicity's sake, we have fitted our data using a model developed for a simplified system with a single pore size, $L$ \cite{Destainville2008}. The fit to the data is imperfect (Fig.~\ref{fgr:figure9}B), most likely because a distribution of pore size is in fact present in our system. Yet, this simple caging model captures the important feature of the MSD (the cross-over), and yields $L = 250$ nm, a value which falls in the expected range of pore diameters for 1.5 wt\% agarose gels. This strongly suggests that temporary caging in the agarose gel pores is what causes the anomalous diffusion behaviour observed for that system. Previous experiments in a comparable agarose system led to the same conclusion \cite{FatinRouge2004,FatinRouge2006}, and comparable caging effects are observed in other systems, such as actin gels \cite{Wong2004}, colloidal suspensions \cite{Weeks2002} and cellular membranes \cite{Murase2004}.

The gel pore size is larger than the diameter of the beads used as probe particles in this study (40~nm) and it is about an order of magnitude smaller than the diameter of the observation volume at the cross-over ($2w_0 \simeq \SI{2}{\micro\meter}$). We are thus detecting the macroscopic regime (i.e. long lengthscales) of the cross-over, where each bead may become trapped multiple times during a passage through the observation volume. At the largest observation volumes, the effect of trapping is absorbed into an effective diffusion coefficient as if the agarose gel were replaced by a homogeneous medium with high effective viscosity. These results emphasize the importance of the relative order of lengthscales of the observation volume, the tracer, and the obstacle mesh when measuring a cross-over effect.

\subsection{The anomalous, yet Brownian motion of proteins in crowded dextran solutions}

The diffusion of proteins in crowded dextran solutions is quite different from the diffusion of beads in agarose gels. We worked at a dextran concentration above the overlap concentration, where the average distance between adjacent dextran molecules ($\simeq 13$ nm) is smaller than their radius of gyration ($\simeq 17$ nm for $276$ kDa dextran molecules \cite{Fundueanu1999,Banks2005}). In these conditions, the polymer chains are entangled, yet fully mobile, as was shown using different techniques \cite{Banks2005,Shakhov2012}. This system (tracer proteins in marginally entangled dextran solutions) has been repeatedly studied by us and others and shown to support anomalous diffusion \cite{Banks2005,Sanabria2007,Goins2008,Szymanski2009,Ernst2012}. Yet explanations as to why differ widely. Using a range of different techniques, Weiss \textit{et al.} \cite{Szymanski2009,Ernst2012} have established that the anomalous diffusion observed in crowded dextran solutions is an ergodic process (ruling out CTRW), and have proposed that it was consistent with FBM. In another series of FCS studies, Waxham \textit{et al.} \cite{Sanabria2007,Goins2008} have argued that the anomalous behaviour observed arose from heterogeneities in the micro-environment experienced by the tracer particles, leading to a discrete distributions of diffusion coefficients. Further, other groups have reported simple diffusion processes in very similar systems, using both FCS \cite{Dauty2004,Dix2008} and pulsed gradient NMR \cite{Shakhov2012}.

Here, directly testing the lengthscale dependence of the diffusion in this system showed that the process is anomalous, as evidenced by the shape of the individual ACFs (Fig.~\ref{fgr:figure3}), with, as for diffusion in the agarose gel, a cross-over from simple to anomalous diffusion below a lengthscale of $\simeq 1 \mu$m, or a timescale of $10$ ms. Further, and in contrast to what was observed for the agarose gels, our data strongly suggests a non-Gaussian diffusion process at short lag times (below $\simeq 10$ ms), for two reasons. First, the values of the anomalous exponent extracted from the ACF and from the diffusion law differ significantly (Fig.~\ref{fgr:figure4}). And second, the apparent MSDs extracted from the ACFs do not overlap at the shorter lag times (Fig.~\ref{fgr:figure5}); if the diffusion process was Gaussian, they should do so. Yet, the diffusion law seems linear in time ($w_0^2 \propto \tau_D$, Fig.~\ref{fgr:figure4}B), indicating that the MSD is also linear.
The fact that mildly entangled fluids can give rise to ``anomalous, yet Brownian'' diffusion, with a non-Gaussian propagator and yet a linear MSD, ties in with recent observations that this type of behaviour is frequent in biological systems \cite{Wang2009,Wang2012}. It remains nevertheless a surprising observation, because a linear MSD is often considered to be associated with a Gaussian propagator, and because most classical anomalous diffusion models predict an MSD deviating from linearity.

The apparently contradicting features we observed for protein diffusion in dextran solutions (namely: \textit{(i)} a linear diffusion law, \textit{(ii)} ACFs deviating from simple diffusion and \textit{(iii)} a propagator switching from Gaussian at large lengthscales to non-Gaussian at shorter lengthscales) can be explained by the ``diffusing diffusivity'' model recently developed by Chubynsky and Slater \cite{Chubynsky2014}. These authors showed that if the lengths of the steps of a particle undergoing a random walk are correlated, but not their direction, then the resulting diffusion is characterized by an MSD linear in time, and a non-Gaussian propagator at lag times shorter than the correlation time of the diffusion coefficient; at longer lag times a Gaussian distribution of displacements is recovered. They proposed that such a process could occur in media that have a spatially varying viscosity. Several independent simulations of tracers in colloidal suspensions have similarly led to the conclusion that spatially heterogeneous crowding could lead to non-Gaussian propagators \cite{Ghosh2016}, and that this non-Gaussianity could be associated with a linear MSD \cite{Kwon2014}. This physical interpretation could apply to crowded dextran solutions, since each dextran molecule forms a finite number of entanglements with its neighbours, and this number may differ from one molecule to the next, resulting in different conformations and mobilities for different dextran molecules. From the point of view of the much smaller tracer molecules, it might result in dynamic micro-environments with different viscosity, as these tracers diffuse across regions spanned by dextran molecules with varying density and mobility. We note that this interpretation of our results is not fundamentally different from that proposed in Refs. \citenum{Sanabria2007} and \citenum{Goins2008}, where it was argued that the anomalous diffusion detected in dextran solutions was arising from the presence of different micro-environments. Similar interpretations have also been invoked in the case of intracellular diffusion: Recently, a spatially varying, random diffusivity \cite{Massignan2014} has been shown to be consistent with receptor motion in dendritic cells \cite{Manzo2015}.

The existence of interactions between the tracer and the larger polymer molecules could provide an alternative physical explanation for the features we observed. Careful analysis of the ACFs (as described in our previous work on this system \cite{Banks2005}) showed that the existence of a prolonged interaction between the streptavidin and the dextran can be ruled out. Indeed, if tracer/polymer complexes are stable enough to diffuse through the observation volume without dissociating, a second separate diffusion term should appear in the ACF, as indeed observed in systems where specific interactions exist between the tracer and the polymer \cite{Vagias2013}. No such term is observed in the case of streptavidin and dextran. This does not immediately exclude, however, a more dynamic interaction scenario, where tracer molecules may bind and unbind during their passage through the detection volume. The duration of the tracer/polymer interaction would then introduce a characteristic timescale in the system, which, if similar to the characteristic residence time of the tracer in the FCS observation volume, might produce a deformation of individual ACFs undistinguishable from that caused by FBM \cite{Vagias2013}. An analytical form for the ACF can be derived in the case of rapid binding and unbinding to a single type of immobile ``traps'' (``stick-and-diffuse'' model, used to explain the mobility of synaptic vesicles \cite{Yeung2007} and transcription factors \cite{Porcher2010}). Examination of the scale-dependence of the ACF predicted by this model shows that in the case of binding to immobile partners, there should be a cross-over in the diffusion law, which is not observed here. However, since dextran molecules are themselves diffusing, it remains possible that transient interactions with a mobile partner may lead to a distribution of diffusivity that fit within the diffusing diffusivity framework, and therefore may also explain our experimental observations.

Importantly, our VLS-FCS experiments finally bridge the lengthscales between diffraction limited FCS experiments, which consistently showed deviations from simple diffusion for proteins in dextran solutions at sub-micron scales \cite{Banks2005,Sanabria2007,Szymanski2009}, and pulsed gradient NMR experiments, which showed that the MSD was linear in time above $1.2~\mu$m in that same system \cite{Shakhov2012}. This bridging had been called for as a means to ``distinguish a cross-over from an inconsistency between methods'' \cite{Saxton2012}. We show here that the apparent discrepancy between these two types of study is indeed simply the result of a cross-over in the diffusion behaviour just below $1~\mu$m, as clearly visible in Fig.~\ref{fgr:figure5}, and as predicted in Ref.~\citenum{Shakhov2012}.

\section{Conclusion}

In conclusion, we have compared the diffusion of tracers in two simple model systems, each of which represent a different form of crowding. Marginally entangled fluids (i.e.\ solutions crowded with entangled yet mobile polymers, here dextrans) give rise to a different type of anomaly in the diffusion than solutions containing a cross-linked polymer network (i.e.\ polymer chains or fibers held together by long-lasting molecular bounds, here agarose gels). This difference is best highlighted in the strikingly different forms of their MSD, as recovered from inversion of the ACFs obtained from FCS experiments. Whereas both show a cross-over towards non-simple diffusion at short lag times, the perfect superposition of the MSDs obtained at different lengthscales in the case of the agarose gel, and their non-superposition in the case of the dextran solution, points towards a Gaussian distribution of displacements in the first case, and a non-Gaussian one in the second case. \\


\section*{Acknowledgments}
This work was funded by the Natural Sciences and Engineering Research Council of Canada (NSERC). DB was the recipient of a CGS-D NSERC scholarship. We thank Dr.\ Thomas Franosch, Dr.\ Gary Slater and Dr.\ Mykyta Chubynsky for helpful discussions.


\footnotesize{

\begin{mcitethebibliography}{105}
\providecommand*{\natexlab}[1]{#1}
\providecommand*{\mciteSetBstSublistMode}[1]{}
\providecommand*{\mciteSetBstMaxWidthForm}[2]{}
\providecommand*{\mciteBstWouldAddEndPuncttrue}
  {\def\EndOfBibitem{\unskip.}}
\providecommand*{\mciteBstWouldAddEndPunctfalse}
  {\let\EndOfBibitem\relax}
\providecommand*{\mciteSetBstMidEndSepPunct}[3]{}
\providecommand*{\mciteSetBstSublistLabelBeginEnd}[3]{}
\providecommand*{\EndOfBibitem}{}
\mciteSetBstSublistMode{f}
\mciteSetBstMaxWidthForm{subitem}
{(\emph{\alph{mcitesubitemcount}})}
\mciteSetBstSublistLabelBeginEnd{\mcitemaxwidthsubitemform\space}
{\relax}{\relax}

\bibitem[Klafter and Sokolov(2005)]{Klafter2005}
J.~Klafter and I.~M. Sokolov, \emph{Phys. World}, 2005, \textbf{18}, 29\relax
\mciteBstWouldAddEndPuncttrue
\mciteSetBstMidEndSepPunct{\mcitedefaultmidpunct}
{\mcitedefaultendpunct}{\mcitedefaultseppunct}\relax
\EndOfBibitem
\bibitem[Luby-Phelps \emph{et~al.}(1987)Luby-Phelps, Castle, Taylor, and
  Lanni]{Luby1987}
K.~Luby-Phelps, P.~E. Castle, D.~L. Taylor and F.~Lanni, \emph{Proc. Nat. Acad.
  Sci. USA}, 1987, \textbf{84}, 4910--4913\relax
\mciteBstWouldAddEndPuncttrue
\mciteSetBstMidEndSepPunct{\mcitedefaultmidpunct}
{\mcitedefaultendpunct}{\mcitedefaultseppunct}\relax
\EndOfBibitem
\bibitem[Kusumi \emph{et~al.}(1993)Kusumi, Sako, and Yamamoto]{Kusumi1993}
A.~Kusumi, Y.~Sako and M.~Yamamoto, \emph{Biophys. J.}, 1993, \textbf{65},
  2021--2040\relax
\mciteBstWouldAddEndPuncttrue
\mciteSetBstMidEndSepPunct{\mcitedefaultmidpunct}
{\mcitedefaultendpunct}{\mcitedefaultseppunct}\relax
\EndOfBibitem
\bibitem[Feder \emph{et~al.}(1996)Feder, Brust-Mascher, Slattery, Baird, and
  Webb]{Feder1996}
T.~J. Feder, I.~Brust-Mascher, J.~P. Slattery, B.~Baird and W.~W. Webb,
  \emph{Biophys. J.}, 1996, \textbf{70}, 2767--2773\relax
\mciteBstWouldAddEndPuncttrue
\mciteSetBstMidEndSepPunct{\mcitedefaultmidpunct}
{\mcitedefaultendpunct}{\mcitedefaultseppunct}\relax
\EndOfBibitem
\bibitem[Wachsmuth \emph{et~al.}(2000)Wachsmuth, Waldeck, and
  Langowski]{Wachsmuth2000}
M.~Wachsmuth, W.~Waldeck and J.~Langowski, \emph{J. Mol. Biol.}, 2000,
  \textbf{298}, 677--689\relax
\mciteBstWouldAddEndPuncttrue
\mciteSetBstMidEndSepPunct{\mcitedefaultmidpunct}
{\mcitedefaultendpunct}{\mcitedefaultseppunct}\relax
\EndOfBibitem
\bibitem[Platani \emph{et~al.}(2002)Platani, Goldberg, Lamond, and
  Swedlow]{Platani2002}
M.~Platani, I.~Goldberg, A.~I. Lamond and J.~R. Swedlow, \emph{Nature Cell
  Biol.}, 2002, \textbf{4}, 502--508\relax
\mciteBstWouldAddEndPuncttrue
\mciteSetBstMidEndSepPunct{\mcitedefaultmidpunct}
{\mcitedefaultendpunct}{\mcitedefaultseppunct}\relax
\EndOfBibitem
\bibitem[Weiss \emph{et~al.}(2003)Weiss, Hashimoto, and Nilsson]{Weiss2003}
M.~Weiss, H.~Hashimoto and T.~Nilsson, \emph{Biophys. J.}, 2003, \textbf{84},
  4043--4052\relax
\mciteBstWouldAddEndPuncttrue
\mciteSetBstMidEndSepPunct{\mcitedefaultmidpunct}
{\mcitedefaultendpunct}{\mcitedefaultseppunct}\relax
\EndOfBibitem
\bibitem[Wawrezinieck \emph{et~al.}(2005)Wawrezinieck, Rigneault, Marguet, and
  Lenne]{Wawrezinieck2005}
L.~Wawrezinieck, H.~Rigneault, D.~Marguet and P.-F. Lenne, \emph{Biophys. J.},
  2005, \textbf{89}, 4029--4042\relax
\mciteBstWouldAddEndPuncttrue
\mciteSetBstMidEndSepPunct{\mcitedefaultmidpunct}
{\mcitedefaultendpunct}{\mcitedefaultseppunct}\relax
\EndOfBibitem
\bibitem[Golding and Cox(2006)]{Golding2006}
I.~Golding and E.~C. Cox, \emph{Phys. Rev. Lett.}, 2006, \textbf{96},
  098102\relax
\mciteBstWouldAddEndPuncttrue
\mciteSetBstMidEndSepPunct{\mcitedefaultmidpunct}
{\mcitedefaultendpunct}{\mcitedefaultseppunct}\relax
\EndOfBibitem
\bibitem[Guigas \emph{et~al.}(2007)Guigas, Kalla, and Weiss]{Guigas2007}
G.~Guigas, C.~Kalla and M.~Weiss, \emph{Biophys. J.}, 2007, \textbf{93},
  316--323\relax
\mciteBstWouldAddEndPuncttrue
\mciteSetBstMidEndSepPunct{\mcitedefaultmidpunct}
{\mcitedefaultendpunct}{\mcitedefaultseppunct}\relax
\EndOfBibitem
\bibitem[Bancaud \emph{et~al.}(2009)Bancaud, Huet, Daigle, Mozziconacci,
  Beaudouin, and Ellenberg]{Bancaud2009}
A.~Bancaud, S.~Huet, N.~Daigle, J.~Mozziconacci, J.~Beaudouin and J.~Ellenberg,
  \emph{EMBO J.}, 2009, \textbf{28}, 3785--3798\relax
\mciteBstWouldAddEndPuncttrue
\mciteSetBstMidEndSepPunct{\mcitedefaultmidpunct}
{\mcitedefaultendpunct}{\mcitedefaultseppunct}\relax
\EndOfBibitem
\bibitem[Abu-Arish \emph{et~al.}(2010)Abu-Arish, Porcher, Czerwonka, Dostatni,
  and Fradin]{Abu2010}
A.~Abu-Arish, A.~Porcher, A.~Czerwonka, N.~Dostatni and C.~Fradin,
  \emph{Biophys. J.}, 2010, \textbf{99}, L33--L35\relax
\mciteBstWouldAddEndPuncttrue
\mciteSetBstMidEndSepPunct{\mcitedefaultmidpunct}
{\mcitedefaultendpunct}{\mcitedefaultseppunct}\relax
\EndOfBibitem
\bibitem[Jeon \emph{et~al.}(2011)Jeon, Tejedor, Burov, Barkai, Selhuber-Unkel,
  Berg-S{\o}rensen, Oddershede, and Metzler]{Jeon2011}
J.-H. Jeon, V.~Tejedor, S.~Burov, E.~Barkai, C.~Selhuber-Unkel,
  K.~Berg-S{\o}rensen, L.~Oddershede and R.~Metzler, \emph{Phys. Rev. Lett.},
  2011, \textbf{106}, 048103\relax
\mciteBstWouldAddEndPuncttrue
\mciteSetBstMidEndSepPunct{\mcitedefaultmidpunct}
{\mcitedefaultendpunct}{\mcitedefaultseppunct}\relax
\EndOfBibitem
\bibitem[Busch \emph{et~al.}(2000)Busch, Kim, and Bloomfield]{Busch2000}
N.~A. Busch, T.~Kim and V.~A. Bloomfield, \emph{Macromolecules}, 2000,
  \textbf{33}, 5932--5937\relax
\mciteBstWouldAddEndPuncttrue
\mciteSetBstMidEndSepPunct{\mcitedefaultmidpunct}
{\mcitedefaultendpunct}{\mcitedefaultseppunct}\relax
\EndOfBibitem
\bibitem[Weiss \emph{et~al.}(2004)Weiss, Elsner, Kartberg, and
  Nilsson]{Weiss2004}
M.~Weiss, M.~Elsner, F.~Kartberg and T.~Nilsson, \emph{Biophys. J.}, 2004,
  \textbf{87}, 3518--3524\relax
\mciteBstWouldAddEndPuncttrue
\mciteSetBstMidEndSepPunct{\mcitedefaultmidpunct}
{\mcitedefaultendpunct}{\mcitedefaultseppunct}\relax
\EndOfBibitem
\bibitem[Banks and Fradin(2005)]{Banks2005}
D.~S. Banks and C.~Fradin, \emph{Biophys. J.}, 2005, \textbf{89},
  2960--2971\relax
\mciteBstWouldAddEndPuncttrue
\mciteSetBstMidEndSepPunct{\mcitedefaultmidpunct}
{\mcitedefaultendpunct}{\mcitedefaultseppunct}\relax
\EndOfBibitem
\bibitem[Amblard \emph{et~al.}(1996)Amblard, Maggs, Yurke, Pargellis, and
  Leibler]{Amblard1996}
F.~Amblard, A.~C. Maggs, B.~Yurke, A.~N. Pargellis and S.~Leibler, \emph{Phys.
  Rev. Lett.}, 1996, \textbf{77}, 4470\relax
\mciteBstWouldAddEndPuncttrue
\mciteSetBstMidEndSepPunct{\mcitedefaultmidpunct}
{\mcitedefaultendpunct}{\mcitedefaultseppunct}\relax
\EndOfBibitem
\bibitem[Valentine \emph{et~al.}(2001)Valentine, Kaplan, Thota, Crocker,
  Gisler, Prud'homme, Beck, and Weitz]{Valentine2001}
M.~T. Valentine, P.~D. Kaplan, D.~Thota, J.~C. Crocker, T.~Gisler, R.~K.
  Prud'homme, M.~Beck and D.~A. Weitz, \emph{Phys. Rev. E}, 2001, \textbf{64},
  061506\relax
\mciteBstWouldAddEndPuncttrue
\mciteSetBstMidEndSepPunct{\mcitedefaultmidpunct}
{\mcitedefaultendpunct}{\mcitedefaultseppunct}\relax
\EndOfBibitem
\bibitem[Wong \emph{et~al.}(2004)Wong, Gardel, Reichman, Weeks, Valentine,
  Bausch, and Weitz]{Wong2004}
I.~Wong, M.~Gardel, D.~Reichman, E.~R. Weeks, M.~Valentine, A.~Bausch and
  D.~Weitz, \emph{Phys. Rev. Lett.}, 2004, \textbf{92}, 178101\relax
\mciteBstWouldAddEndPuncttrue
\mciteSetBstMidEndSepPunct{\mcitedefaultmidpunct}
{\mcitedefaultendpunct}{\mcitedefaultseppunct}\relax
\EndOfBibitem
\bibitem[Fatin-Rouge \emph{et~al.}(2004)Fatin-Rouge, Starchev, and
  Buffle]{FatinRouge2004}
N.~Fatin-Rouge, K.~Starchev and J.~Buffle, \emph{Biophys. J.}, 2004,
  \textbf{86}, 2710--2719\relax
\mciteBstWouldAddEndPuncttrue
\mciteSetBstMidEndSepPunct{\mcitedefaultmidpunct}
{\mcitedefaultendpunct}{\mcitedefaultseppunct}\relax
\EndOfBibitem
\bibitem[Weeks and Weitz(2002)]{Weeks2002}
E.~R. Weeks and D.~Weitz, \emph{Chem. Phys.}, 2002, \textbf{284},
  361--367\relax
\mciteBstWouldAddEndPuncttrue
\mciteSetBstMidEndSepPunct{\mcitedefaultmidpunct}
{\mcitedefaultendpunct}{\mcitedefaultseppunct}\relax
\EndOfBibitem
\bibitem[Lenne \emph{et~al.}(2006)Lenne, Wawrezinieck, Conchonaud, Wurtz,
  Boned, Guo, Rigneault, He, and Marguet]{Lenne2006}
P.-F. Lenne, L.~Wawrezinieck, F.~Conchonaud, O.~Wurtz, A.~Boned, X.-J. Guo,
  H.~Rigneault, H.-T. He and D.~Marguet, \emph{EMBO J.}, 2006, \textbf{25},
  3245--3256\relax
\mciteBstWouldAddEndPuncttrue
\mciteSetBstMidEndSepPunct{\mcitedefaultmidpunct}
{\mcitedefaultendpunct}{\mcitedefaultseppunct}\relax
\EndOfBibitem
\bibitem[Saxton(1994)]{Saxton1994}
M.~J. Saxton, \emph{Biophys. J.}, 1994, \textbf{66}, 394--401\relax
\mciteBstWouldAddEndPuncttrue
\mciteSetBstMidEndSepPunct{\mcitedefaultmidpunct}
{\mcitedefaultendpunct}{\mcitedefaultseppunct}\relax
\EndOfBibitem
\bibitem[Szymanski and Weiss(2009)]{Szymanski2009}
J.~Szymanski and M.~Weiss, \emph{Phys. Rev. Lett.}, 2009, \textbf{103},
  038102\relax
\mciteBstWouldAddEndPuncttrue
\mciteSetBstMidEndSepPunct{\mcitedefaultmidpunct}
{\mcitedefaultendpunct}{\mcitedefaultseppunct}\relax
\EndOfBibitem
\bibitem[Saxton(1996)]{Saxton1996}
M.~J. Saxton, \emph{Biophys. J.}, 1996, \textbf{70}, 1250--1262\relax
\mciteBstWouldAddEndPuncttrue
\mciteSetBstMidEndSepPunct{\mcitedefaultmidpunct}
{\mcitedefaultendpunct}{\mcitedefaultseppunct}\relax
\EndOfBibitem
\bibitem[Bronstein \emph{et~al.}(2009)Bronstein, Israel, Kepten, Mai, Shav-Tal,
  Barkai, and Garini]{Bronstein2009}
I.~Bronstein, Y.~Israel, E.~Kepten, S.~Mai, Y.~Shav-Tal, E.~Barkai and
  Y.~Garini, \emph{Phys. Rev. Lett.}, 2009, \textbf{103}, 018102\relax
\mciteBstWouldAddEndPuncttrue
\mciteSetBstMidEndSepPunct{\mcitedefaultmidpunct}
{\mcitedefaultendpunct}{\mcitedefaultseppunct}\relax
\EndOfBibitem
\bibitem[Metzler and Klafter(2000)]{Metzler2000}
R.~Metzler and J.~Klafter, \emph{Phys. Rep.}, 2000, \textbf{339}, 1--77\relax
\mciteBstWouldAddEndPuncttrue
\mciteSetBstMidEndSepPunct{\mcitedefaultmidpunct}
{\mcitedefaultendpunct}{\mcitedefaultseppunct}\relax
\EndOfBibitem
\bibitem[Sokolov(2012)]{Sokolov2012}
I.~M. Sokolov, \emph{Soft Matter}, 2012, \textbf{8}, 9043--9052\relax
\mciteBstWouldAddEndPuncttrue
\mciteSetBstMidEndSepPunct{\mcitedefaultmidpunct}
{\mcitedefaultendpunct}{\mcitedefaultseppunct}\relax
\EndOfBibitem
\bibitem[H{\"o}fling and Franosch(2013)]{Hofling2013}
F.~H{\"o}fling and T.~Franosch, \emph{Rep. Prog. Phys.}, 2013, \textbf{76},
  046602\relax
\mciteBstWouldAddEndPuncttrue
\mciteSetBstMidEndSepPunct{\mcitedefaultmidpunct}
{\mcitedefaultendpunct}{\mcitedefaultseppunct}\relax
\EndOfBibitem
\bibitem[Bressloff and Newby(2013)]{Bressloff2013}
P.~C. Bressloff and J.~M. Newby, \emph{Rev. Mod. Phys.}, 2013, \textbf{85},
  135--196\relax
\mciteBstWouldAddEndPuncttrue
\mciteSetBstMidEndSepPunct{\mcitedefaultmidpunct}
{\mcitedefaultendpunct}{\mcitedefaultseppunct}\relax
\EndOfBibitem
\bibitem[Metzler \emph{et~al.}(2014)Metzler, Jeon, Cherstvy, and
  Barkai]{Metzler2014}
R.~Metzler, J.-H. Jeon, A.~G. Cherstvy and E.~Barkai, \emph{Phys. Chem. Chem.
  Phys.}, 2014, \textbf{16}, 24128--24164\relax
\mciteBstWouldAddEndPuncttrue
\mciteSetBstMidEndSepPunct{\mcitedefaultmidpunct}
{\mcitedefaultendpunct}{\mcitedefaultseppunct}\relax
\EndOfBibitem
\bibitem[Weiss(2014)]{Weiss2014}
M.~Weiss, \emph{New Models of the Cell Nucleus: Crowding, Entropic Forces,
  Phase Separation, and Fractals}, Academic Press, 2014, vol. 307, ch.~11, pp.
  383--417\relax
\mciteBstWouldAddEndPuncttrue
\mciteSetBstMidEndSepPunct{\mcitedefaultmidpunct}
{\mcitedefaultendpunct}{\mcitedefaultseppunct}\relax
\EndOfBibitem
\bibitem[Meroz and Sokolov(2015)]{Meroz2015}
Y.~Meroz and I.~M. Sokolov, \emph{Phys. Rep.}, 2015, \textbf{573}, 1--29\relax
\mciteBstWouldAddEndPuncttrue
\mciteSetBstMidEndSepPunct{\mcitedefaultmidpunct}
{\mcitedefaultendpunct}{\mcitedefaultseppunct}\relax
\EndOfBibitem
\bibitem[Yeung \emph{et~al.}(2007)Yeung, Shtrahman, and Wu]{Yeung2007}
C.~Yeung, M.~Shtrahman and X.-l. Wu, \emph{Biophys. J.}, 2007, \textbf{92},
  2271--2280\relax
\mciteBstWouldAddEndPuncttrue
\mciteSetBstMidEndSepPunct{\mcitedefaultmidpunct}
{\mcitedefaultendpunct}{\mcitedefaultseppunct}\relax
\EndOfBibitem
\bibitem[Fujiwara \emph{et~al.}(2002)Fujiwara, Ritchie, Murakoshi, Jacobson,
  and Kusumi]{Fujiwara2002}
T.~Fujiwara, K.~Ritchie, H.~Murakoshi, K.~Jacobson and A.~Kusumi, \emph{J. Cell
  Biol.}, 2002, \textbf{157}, 1071--1082\relax
\mciteBstWouldAddEndPuncttrue
\mciteSetBstMidEndSepPunct{\mcitedefaultmidpunct}
{\mcitedefaultendpunct}{\mcitedefaultseppunct}\relax
\EndOfBibitem
\bibitem[Waharte \emph{et~al.}(2005)Waharte, Brown, Coscoy, Coudrier, and
  Amblard]{Waharte2005}
F.~Waharte, C.~M. Brown, S.~Coscoy, E.~Coudrier and F.~Amblard, \emph{Biophys.
  J.}, 2005, \textbf{88}, 1467--1478\relax
\mciteBstWouldAddEndPuncttrue
\mciteSetBstMidEndSepPunct{\mcitedefaultmidpunct}
{\mcitedefaultendpunct}{\mcitedefaultseppunct}\relax
\EndOfBibitem
\bibitem[Schwille \emph{et~al.}(1999)Schwille, Korlach, and Webb]{Schwille1999}
P.~Schwille, J.~Korlach and W.~W. Webb, \emph{Cytometry}, 1999, \textbf{36},
  176--182\relax
\mciteBstWouldAddEndPuncttrue
\mciteSetBstMidEndSepPunct{\mcitedefaultmidpunct}
{\mcitedefaultendpunct}{\mcitedefaultseppunct}\relax
\EndOfBibitem
\bibitem[Wu and Berland(2008)]{Wu2008}
J.~Wu and K.~M. Berland, \emph{Biophys. J.}, 2008, \textbf{95},
  2049--2052\relax
\mciteBstWouldAddEndPuncttrue
\mciteSetBstMidEndSepPunct{\mcitedefaultmidpunct}
{\mcitedefaultendpunct}{\mcitedefaultseppunct}\relax
\EndOfBibitem
\bibitem[Shusterman \emph{et~al.}(2004)Shusterman, Alon, Gavrinyov, and
  Krichevsky]{Shusterman2004}
R.~Shusterman, S.~Alon, T.~Gavrinyov and O.~Krichevsky, \emph{Phys. Rev.
  Lett.}, 2004, \textbf{92}, 048303\relax
\mciteBstWouldAddEndPuncttrue
\mciteSetBstMidEndSepPunct{\mcitedefaultmidpunct}
{\mcitedefaultendpunct}{\mcitedefaultseppunct}\relax
\EndOfBibitem
\bibitem[Shusterman \emph{et~al.}(2008)Shusterman, Gavrinyov, and
  Krichevsky]{Shusterman2008}
R.~Shusterman, T.~Gavrinyov and O.~Krichevsky, \emph{Phys. Rev. Lett.}, 2008,
  \textbf{100}, 098102\relax
\mciteBstWouldAddEndPuncttrue
\mciteSetBstMidEndSepPunct{\mcitedefaultmidpunct}
{\mcitedefaultendpunct}{\mcitedefaultseppunct}\relax
\EndOfBibitem
\bibitem[Horton \emph{et~al.}(2010)Horton, H{\"o}fling, R{\"a}dler, and
  Franosch]{Horton2010}
M.~R. Horton, F.~H{\"o}fling, J.~O. R{\"a}dler and T.~Franosch, \emph{Soft
  Matter}, 2010, \textbf{6}, 2648--2656\relax
\mciteBstWouldAddEndPuncttrue
\mciteSetBstMidEndSepPunct{\mcitedefaultmidpunct}
{\mcitedefaultendpunct}{\mcitedefaultseppunct}\relax
\EndOfBibitem
\bibitem[Wawrezinieck \emph{et~al.}(2004)Wawrezinieck, Lenne, Marguet, and
  Rigneault]{Wawrezinieck2004}
L.~Wawrezinieck, P.-F. Lenne, D.~Marguet and H.~Rigneault, Photonics Europe,
  2004, pp. 92--102\relax
\mciteBstWouldAddEndPuncttrue
\mciteSetBstMidEndSepPunct{\mcitedefaultmidpunct}
{\mcitedefaultendpunct}{\mcitedefaultseppunct}\relax
\EndOfBibitem
\bibitem[Masuda \emph{et~al.}(2005)Masuda, Ushida, and Okamoto]{Masuda2005}
A.~Masuda, K.~Ushida and T.~Okamoto, \emph{Biophys. J.}, 2005, \textbf{88},
  3584--3591\relax
\mciteBstWouldAddEndPuncttrue
\mciteSetBstMidEndSepPunct{\mcitedefaultmidpunct}
{\mcitedefaultendpunct}{\mcitedefaultseppunct}\relax
\EndOfBibitem
\bibitem[Eggeling \emph{et~al.}(2008)Eggeling, Ringemann, Medda, Schwarzmann,
  Sandhoff, Polyakova, Belov, Hein, von Middendorff,
  Sch{\"o}nle,\emph{et~al.}]{Eggeling2008}
C.~Eggeling, C.~Ringemann, R.~Medda, G.~Schwarzmann, K.~Sandhoff, S.~Polyakova,
  V.~N. Belov, B.~Hein, C.~von Middendorff, A.~Sch{\"o}nle \emph{et~al.},
  \emph{Nature}, 2008, \textbf{457}, 1159--1162\relax
\mciteBstWouldAddEndPuncttrue
\mciteSetBstMidEndSepPunct{\mcitedefaultmidpunct}
{\mcitedefaultendpunct}{\mcitedefaultseppunct}\relax
\EndOfBibitem
\bibitem[H{\"o}fling \emph{et~al.}(2011)H{\"o}fling, Bamberg, and
  Franosch]{Hofling2011}
F.~H{\"o}fling, K.-U. Bamberg and T.~Franosch, \emph{Soft Matter}, 2011,
  \textbf{7}, 1358--1363\relax
\mciteBstWouldAddEndPuncttrue
\mciteSetBstMidEndSepPunct{\mcitedefaultmidpunct}
{\mcitedefaultendpunct}{\mcitedefaultseppunct}\relax
\EndOfBibitem
\bibitem[Favard \emph{et~al.}(2011)Favard, Wenger, Lenne, and
  Rigneault]{Favard2011}
C.~Favard, J.~Wenger, P.-F. Lenne and H.~Rigneault, \emph{Biophys. J.}, 2011,
  \textbf{100}, 1242--1251\relax
\mciteBstWouldAddEndPuncttrue
\mciteSetBstMidEndSepPunct{\mcitedefaultmidpunct}
{\mcitedefaultendpunct}{\mcitedefaultseppunct}\relax
\EndOfBibitem
\bibitem[Masuda \emph{et~al.}(2005)Masuda, Ushida, and Okamoto]{Masuda2005b}
A.~Masuda, K.~Ushida and T.~Okamoto, \emph{Phys. Rev. E}, 2005, \textbf{72},
  060101\relax
\mciteBstWouldAddEndPuncttrue
\mciteSetBstMidEndSepPunct{\mcitedefaultmidpunct}
{\mcitedefaultendpunct}{\mcitedefaultseppunct}\relax
\EndOfBibitem
\bibitem[Masuda \emph{et~al.}(2006)Masuda, Ushida, and Okamoto]{Masuda2006}
A.~Masuda, K.~Ushida and T.~Okamoto, \emph{J. Photochem. Photobiol. A}, 2006,
  \textbf{183}, 304--308\relax
\mciteBstWouldAddEndPuncttrue
\mciteSetBstMidEndSepPunct{\mcitedefaultmidpunct}
{\mcitedefaultendpunct}{\mcitedefaultseppunct}\relax
\EndOfBibitem
\bibitem[King \emph{et~al.}(2014)King, Yu, Wilson, and Granick]{King2014}
J.~T. King, C.~Yu, W.~L. Wilson and S.~Granick, \emph{ACS nano}, 2014,
  \textbf{8}, 8802--8809\relax
\mciteBstWouldAddEndPuncttrue
\mciteSetBstMidEndSepPunct{\mcitedefaultmidpunct}
{\mcitedefaultendpunct}{\mcitedefaultseppunct}\relax
\EndOfBibitem
\bibitem[Sanabria \emph{et~al.}(2007)Sanabria, Kubota, and
  Waxham]{Sanabria2007}
H.~Sanabria, Y.~Kubota and M.~N. Waxham, \emph{Biophys. J.}, 2007, \textbf{92},
  313--322\relax
\mciteBstWouldAddEndPuncttrue
\mciteSetBstMidEndSepPunct{\mcitedefaultmidpunct}
{\mcitedefaultendpunct}{\mcitedefaultseppunct}\relax
\EndOfBibitem
\bibitem[Goins \emph{et~al.}(2008)Goins, Sanabria, and Waxham]{Goins2008}
A.~B. Goins, H.~Sanabria and M.~N. Waxham, \emph{Biophys. J.}, 2008,
  \textbf{95}, 5362--5373\relax
\mciteBstWouldAddEndPuncttrue
\mciteSetBstMidEndSepPunct{\mcitedefaultmidpunct}
{\mcitedefaultendpunct}{\mcitedefaultseppunct}\relax
\EndOfBibitem
\bibitem[Pan \emph{et~al.}(2009)Pan, Filobelo, Pham, Galkin, Uzunova, and
  Vekilov]{Pan2009}
W.~Pan, L.~Filobelo, N.~D. Pham, O.~Galkin, V.~V. Uzunova and P.~G. Vekilov,
  \emph{Phys. Rev. Lett.}, 2009, \textbf{102}, 058101\relax
\mciteBstWouldAddEndPuncttrue
\mciteSetBstMidEndSepPunct{\mcitedefaultmidpunct}
{\mcitedefaultendpunct}{\mcitedefaultseppunct}\relax
\EndOfBibitem
\bibitem[Ernst \emph{et~al.}(2012)Ernst, Hellmann, K{\"o}hler, and
  Weiss]{Ernst2012}
D.~Ernst, M.~Hellmann, J.~K{\"o}hler and M.~Weiss, \emph{Soft Matter}, 2012,
  \textbf{8}, 4886--4889\relax
\mciteBstWouldAddEndPuncttrue
\mciteSetBstMidEndSepPunct{\mcitedefaultmidpunct}
{\mcitedefaultendpunct}{\mcitedefaultseppunct}\relax
\EndOfBibitem
\bibitem[Dauty and Verkman(2004)]{Dauty2004}
E.~Dauty and A.~Verkman, \emph{J. Mol. Recog.}, 2004, \textbf{17},
  441--447\relax
\mciteBstWouldAddEndPuncttrue
\mciteSetBstMidEndSepPunct{\mcitedefaultmidpunct}
{\mcitedefaultendpunct}{\mcitedefaultseppunct}\relax
\EndOfBibitem
\bibitem[Dix and Verkman(2008)]{Dix2008}
J.~A. Dix and A.~Verkman, \emph{Annu. Rev. Biophys.}, 2008, \textbf{37},
  247--263\relax
\mciteBstWouldAddEndPuncttrue
\mciteSetBstMidEndSepPunct{\mcitedefaultmidpunct}
{\mcitedefaultendpunct}{\mcitedefaultseppunct}\relax
\EndOfBibitem
\bibitem[Shakhov \emph{et~al.}(2012)Shakhov, Valiullin, and
  K\"arger]{Shakhov2012}
A.~Shakhov, R.~Valiullin and J.~K\"arger, \emph{J. Phys. Chem. Lett.}, 2012,
  \textbf{3}, 1854--1857\relax
\mciteBstWouldAddEndPuncttrue
\mciteSetBstMidEndSepPunct{\mcitedefaultmidpunct}
{\mcitedefaultendpunct}{\mcitedefaultseppunct}\relax
\EndOfBibitem
\bibitem[Lead \emph{et~al.}(2003)Lead, Starchev, and Wilkinson]{Lead2003}
J.~Lead, K.~Starchev and K.~Wilkinson, \emph{Environ. Sci. Technol.}, 2003,
  \textbf{37}, 482--487\relax
\mciteBstWouldAddEndPuncttrue
\mciteSetBstMidEndSepPunct{\mcitedefaultmidpunct}
{\mcitedefaultendpunct}{\mcitedefaultseppunct}\relax
\EndOfBibitem
\bibitem[Fatin-Rouge \emph{et~al.}(2006)Fatin-Rouge, Wilkinson, and
  Buffle]{FatinRouge2006}
N.~Fatin-Rouge, K.~J. Wilkinson and J.~Buffle, \emph{J. Phys. Chem. B}, 2006,
  \textbf{110}, 20133--20142\relax
\mciteBstWouldAddEndPuncttrue
\mciteSetBstMidEndSepPunct{\mcitedefaultmidpunct}
{\mcitedefaultendpunct}{\mcitedefaultseppunct}\relax
\EndOfBibitem
\bibitem[Labille \emph{et~al.}(2007)Labille, Fatin-Rouge, and
  Buffle]{Labille2007}
J.~Labille, N.~Fatin-Rouge and J.~Buffle, \emph{Langmuir}, 2007, \textbf{23},
  2083--2090\relax
\mciteBstWouldAddEndPuncttrue
\mciteSetBstMidEndSepPunct{\mcitedefaultmidpunct}
{\mcitedefaultendpunct}{\mcitedefaultseppunct}\relax
\EndOfBibitem
\bibitem[Mandelbrot and Van~Ness(1968)]{Mandelbrot1968}
B.~B. Mandelbrot and J.~W. Van~Ness, \emph{SIAM review}, 1968, \textbf{10},
  422--437\relax
\mciteBstWouldAddEndPuncttrue
\mciteSetBstMidEndSepPunct{\mcitedefaultmidpunct}
{\mcitedefaultendpunct}{\mcitedefaultseppunct}\relax
\EndOfBibitem
\bibitem[Goychuk(2009)]{Goychuk2009}
I.~Goychuk, \emph{Phys. Rev. E}, 2009, \textbf{80}, 046125\relax
\mciteBstWouldAddEndPuncttrue
\mciteSetBstMidEndSepPunct{\mcitedefaultmidpunct}
{\mcitedefaultendpunct}{\mcitedefaultseppunct}\relax
\EndOfBibitem
\bibitem[Hess and Webb(2002)]{Hess2002}
S.~T. Hess and W.~W. Webb, \emph{Biophys. J.}, 2002, \textbf{83},
  2300--2317\relax
\mciteBstWouldAddEndPuncttrue
\mciteSetBstMidEndSepPunct{\mcitedefaultmidpunct}
{\mcitedefaultendpunct}{\mcitedefaultseppunct}\relax
\EndOfBibitem
\bibitem[Culbertson \emph{et~al.}(2002)Culbertson, Jacobson, and
  Michael~Ramsey]{Culbertson2002}
C.~T. Culbertson, S.~C. Jacobson and J.~Michael~Ramsey, \emph{Talanta}, 2002,
  \textbf{56}, 365--373\relax
\mciteBstWouldAddEndPuncttrue
\mciteSetBstMidEndSepPunct{\mcitedefaultmidpunct}
{\mcitedefaultendpunct}{\mcitedefaultseppunct}\relax
\EndOfBibitem
\bibitem[Petr{\'a}{\v{s}}ek and Schwille(2008)]{Petrasek2008}
Z.~Petr{\'a}{\v{s}}ek and P.~Schwille, \emph{Biophys. J.}, 2008, \textbf{94},
  1437--1448\relax
\mciteBstWouldAddEndPuncttrue
\mciteSetBstMidEndSepPunct{\mcitedefaultmidpunct}
{\mcitedefaultendpunct}{\mcitedefaultseppunct}\relax
\EndOfBibitem
\bibitem[Widengren \emph{et~al.}(1995)Widengren, Mets, and
  Rigler]{Widengren1995}
J.~Widengren, U.~Mets and R.~Rigler, \emph{J. Phys. Chem.}, 1995, \textbf{99},
  13368--13379\relax
\mciteBstWouldAddEndPuncttrue
\mciteSetBstMidEndSepPunct{\mcitedefaultmidpunct}
{\mcitedefaultendpunct}{\mcitedefaultseppunct}\relax
\EndOfBibitem
\bibitem[Buschmann \emph{et~al.}(2003)Buschmann, Weston, and
  Sauer]{Buschmann2003}
V.~Buschmann, K.~D. Weston and M.~Sauer, \emph{Bioconjugate Chem.}, 2003,
  \textbf{14}, 195--204\relax
\mciteBstWouldAddEndPuncttrue
\mciteSetBstMidEndSepPunct{\mcitedefaultmidpunct}
{\mcitedefaultendpunct}{\mcitedefaultseppunct}\relax
\EndOfBibitem
\bibitem[Mazouchi \emph{et~al.}(2013)Mazouchi, Bahram, and
  Gradinaru]{Mazouchi2013}
A.~Mazouchi, A.~Bahram and C.~C. Gradinaru, \emph{J. Phys. Chem. B}, 2013,
  \textbf{117}, 11100--11111\relax
\mciteBstWouldAddEndPuncttrue
\mciteSetBstMidEndSepPunct{\mcitedefaultmidpunct}
{\mcitedefaultendpunct}{\mcitedefaultseppunct}\relax
\EndOfBibitem
\bibitem[Brown \emph{et~al.}(1999)Brown, Wu, Zipfel, and Webb]{Brown1999}
E.~B. Brown, E.~S. Wu, W.~Zipfel and W.~W. Webb, \emph{Biophys. J.}, 1999,
  \textbf{77}, 2837--2849\relax
\mciteBstWouldAddEndPuncttrue
\mciteSetBstMidEndSepPunct{\mcitedefaultmidpunct}
{\mcitedefaultendpunct}{\mcitedefaultseppunct}\relax
\EndOfBibitem
\bibitem[Lubelski and Klafter(2009)]{Lubelski2009}
A.~Lubelski and J.~Klafter, \emph{Biophys. J.}, 2009, \textbf{96},
  2055--2063\relax
\mciteBstWouldAddEndPuncttrue
\mciteSetBstMidEndSepPunct{\mcitedefaultmidpunct}
{\mcitedefaultendpunct}{\mcitedefaultseppunct}\relax
\EndOfBibitem
\bibitem[Humpol{\'\i}{\v{c}}kov{\'a}
  \emph{et~al.}(2006)Humpol{\'\i}{\v{c}}kov{\'a}, Gielen, Benda, Fagulova,
  Vercammen, VandeVen, Hof, Ameloot, and Engelborghs]{Humpolivckova2006}
J.~Humpol{\'\i}{\v{c}}kov{\'a}, E.~Gielen, A.~Benda, V.~Fagulova, J.~Vercammen,
  M.~VandeVen, M.~Hof, M.~Ameloot and Y.~Engelborghs, \emph{Biophys. J.}, 2006,
  \textbf{91}, L23--L25\relax
\mciteBstWouldAddEndPuncttrue
\mciteSetBstMidEndSepPunct{\mcitedefaultmidpunct}
{\mcitedefaultendpunct}{\mcitedefaultseppunct}\relax
\EndOfBibitem
\bibitem[Bag \emph{et~al.}(2012)Bag, Sankaran, Paul, Kraut, and
  Wohland]{Bag2012}
N.~Bag, J.~Sankaran, A.~Paul, R.~S. Kraut and T.~Wohland, \emph{ChemPhysChem},
  2012, \textbf{13}, 2784--2794\relax
\mciteBstWouldAddEndPuncttrue
\mciteSetBstMidEndSepPunct{\mcitedefaultmidpunct}
{\mcitedefaultendpunct}{\mcitedefaultseppunct}\relax
\EndOfBibitem
\bibitem[Di~Rienzo \emph{et~al.}(2013)Di~Rienzo, Gratton, Beltram, and
  Cardarelli]{Di2013}
C.~Di~Rienzo, E.~Gratton, F.~Beltram and F.~Cardarelli, \emph{Proc. Nat. Acad.
  Sci. USA}, 2013, \textbf{110}, 12307--12312\relax
\mciteBstWouldAddEndPuncttrue
\mciteSetBstMidEndSepPunct{\mcitedefaultmidpunct}
{\mcitedefaultendpunct}{\mcitedefaultseppunct}\relax
\EndOfBibitem
\bibitem[Di~Rienzo \emph{et~al.}(2014)Di~Rienzo, Piazza, Gratton, Beltram, and
  Cardarelli]{Di2014}
C.~Di~Rienzo, V.~Piazza, E.~Gratton, F.~Beltram and F.~Cardarelli, \emph{Nat.
  Commun.}, 2014, \textbf{5}, 1--8\relax
\mciteBstWouldAddEndPuncttrue
\mciteSetBstMidEndSepPunct{\mcitedefaultmidpunct}
{\mcitedefaultendpunct}{\mcitedefaultseppunct}\relax
\EndOfBibitem
\bibitem[Kastrup \emph{et~al.}(2005)Kastrup, Blom, Eggeling, and
  Hell]{Kastrup2005}
L.~Kastrup, H.~Blom, C.~Eggeling and S.~W. Hell, \emph{Phys. Rev. Lett.}, 2005,
  \textbf{94}, 178104\relax
\mciteBstWouldAddEndPuncttrue
\mciteSetBstMidEndSepPunct{\mcitedefaultmidpunct}
{\mcitedefaultendpunct}{\mcitedefaultseppunct}\relax
\EndOfBibitem
\bibitem[Mueller \emph{et~al.}(2012)Mueller, Honigmann, Ringemann, Medda,
  Schwarzmann, and Eggeling]{Mueller2012}
V.~Mueller, A.~Honigmann, C.~Ringemann, R.~Medda, G.~Schwarzmann and
  C.~Eggeling, \emph{Methods Enzymol.}, 2012, \textbf{519}, 1--38\relax
\mciteBstWouldAddEndPuncttrue
\mciteSetBstMidEndSepPunct{\mcitedefaultmidpunct}
{\mcitedefaultendpunct}{\mcitedefaultseppunct}\relax
\EndOfBibitem
\bibitem[Samiee \emph{et~al.}(2006)Samiee, Moran-Mirabal, Cheung, and
  Craighead]{Samiee2006}
K.~Samiee, J.~Moran-Mirabal, Y.~Cheung and H.~Craighead, \emph{Biophys. J.},
  2006, \textbf{90}, 3288--3299\relax
\mciteBstWouldAddEndPuncttrue
\mciteSetBstMidEndSepPunct{\mcitedefaultmidpunct}
{\mcitedefaultendpunct}{\mcitedefaultseppunct}\relax
\EndOfBibitem
\bibitem[Vobornik \emph{et~al.}(2008)Vobornik, Banks, Lu, Fradin, Taylor, and
  Johnston]{Vobornik2008}
D.~Vobornik, D.~S. Banks, Z.~Lu, C.~Fradin, R.~Taylor and L.~J. Johnston,
  \emph{Appl. Phys. Lett.}, 2008, \textbf{93}, 163904\relax
\mciteBstWouldAddEndPuncttrue
\mciteSetBstMidEndSepPunct{\mcitedefaultmidpunct}
{\mcitedefaultendpunct}{\mcitedefaultseppunct}\relax
\EndOfBibitem
\bibitem[Manzo \emph{et~al.}(2011)Manzo, van Zanten, and
  Garcia-Parajo]{Manzo2011}
C.~Manzo, T.~S. van Zanten and M.~F. Garcia-Parajo, \emph{Biophys. J.}, 2011,
  \textbf{100}, L8--L10\relax
\mciteBstWouldAddEndPuncttrue
\mciteSetBstMidEndSepPunct{\mcitedefaultmidpunct}
{\mcitedefaultendpunct}{\mcitedefaultseppunct}\relax
\EndOfBibitem
\bibitem[Thorneywork \emph{et~al.}()Thorneywork, Aarts, Horbach, and
  Dullens]{Thorneywork2016}
A.~L. Thorneywork, D.~G. A.~L. Aarts, J.~Horbach and R.~P.~A. Dullens,
  \emph{under review}\relax
\mciteBstWouldAddEndPuncttrue
\mciteSetBstMidEndSepPunct{\mcitedefaultmidpunct}
{\mcitedefaultendpunct}{\mcitedefaultseppunct}\relax
\EndOfBibitem
\bibitem[Kwon \emph{et~al.}(2014)Kwon, Sung, and Yethiraj]{Kwon2014}
G.~Kwon, B.~J. Sung and A.~Yethiraj, \emph{J. Phys. Chem. B}, 2014,
  \textbf{118}, 8128--8134\relax
\mciteBstWouldAddEndPuncttrue
\mciteSetBstMidEndSepPunct{\mcitedefaultmidpunct}
{\mcitedefaultendpunct}{\mcitedefaultseppunct}\relax
\EndOfBibitem
\bibitem[Phillies(2015)]{Phillies2015}
G.~D.~J. Phillies, \emph{Soft matter}, 2015, \textbf{11}, 580--586\relax
\mciteBstWouldAddEndPuncttrue
\mciteSetBstMidEndSepPunct{\mcitedefaultmidpunct}
{\mcitedefaultendpunct}{\mcitedefaultseppunct}\relax
\EndOfBibitem
\bibitem[Ghosh \emph{et~al.}(2016)Ghosh, Cherstvy, Grebenkov, and
  Metzler]{Ghosh2016}
S.~K. Ghosh, A.~G. Cherstvy, D.~S. Grebenkov and R.~Metzler, \emph{New J.
  Phys.}, 2016, \textbf{18}, 013027\relax
\mciteBstWouldAddEndPuncttrue
\mciteSetBstMidEndSepPunct{\mcitedefaultmidpunct}
{\mcitedefaultendpunct}{\mcitedefaultseppunct}\relax
\EndOfBibitem
\bibitem[Wang \emph{et~al.}(2009)Wang, Anthony, Bae, and Granick]{Wang2009}
B.~Wang, S.~M. Anthony, S.~C. Bae and S.~Granick, \emph{Proc. Nat. Acad. Sci.
  USA}, 2009, \textbf{106}, 15160--15164\relax
\mciteBstWouldAddEndPuncttrue
\mciteSetBstMidEndSepPunct{\mcitedefaultmidpunct}
{\mcitedefaultendpunct}{\mcitedefaultseppunct}\relax
\EndOfBibitem
\bibitem[Wang \emph{et~al.}(2012)Wang, Kuo, Bae, and Granick]{Wang2012}
B.~Wang, J.~Kuo, S.~C. Bae and S.~Granick, \emph{Nat. Materials}, 2012,
  \textbf{11}, 481--485\relax
\mciteBstWouldAddEndPuncttrue
\mciteSetBstMidEndSepPunct{\mcitedefaultmidpunct}
{\mcitedefaultendpunct}{\mcitedefaultseppunct}\relax
\EndOfBibitem
\bibitem[Xue \emph{et~al.}(2016)Xue, Zheng, Chen, Tian, and Hu]{Xue2016}
C.~Xue, X.~Zheng, K.~Chen, Y.~Tian and G.~Hu, \emph{J. Phys. Chem. Lett.},
  2016, \textbf{7}, 514--519\relax
\mciteBstWouldAddEndPuncttrue
\mciteSetBstMidEndSepPunct{\mcitedefaultmidpunct}
{\mcitedefaultendpunct}{\mcitedefaultseppunct}\relax
\EndOfBibitem
\bibitem[Sebastian(1995)]{Sebastian1995}
K.~Sebastian, \emph{J. Phys. A}, 1995, \textbf{28}, 4305\relax
\mciteBstWouldAddEndPuncttrue
\mciteSetBstMidEndSepPunct{\mcitedefaultmidpunct}
{\mcitedefaultendpunct}{\mcitedefaultseppunct}\relax
\EndOfBibitem
\bibitem[Klafter and Silbey(1980)]{Klafter1980}
J.~Klafter and R.~Silbey, \emph{Phys. Rev. Lett.}, 1980, \textbf{44}, 55\relax
\mciteBstWouldAddEndPuncttrue
\mciteSetBstMidEndSepPunct{\mcitedefaultmidpunct}
{\mcitedefaultendpunct}{\mcitedefaultseppunct}\relax
\EndOfBibitem
\bibitem[Havlin and Ben-Avraham(1987)]{Havlin1987}
S.~Havlin and D.~Ben-Avraham, \emph{Adv. Phys.}, 1987, \textbf{36},
  695--798\relax
\mciteBstWouldAddEndPuncttrue
\mciteSetBstMidEndSepPunct{\mcitedefaultmidpunct}
{\mcitedefaultendpunct}{\mcitedefaultseppunct}\relax
\EndOfBibitem
\bibitem[H{\"o}f{}ling \emph{et~al.}(2006)H{\"o}f{}ling, Franosch, and
  Frey]{Hofling2006}
F.~H{\"o}f{}ling, T.~Franosch and E.~Frey, \emph{Phys. Rev. Lett.}, 2006,
  \textbf{96}, 165901\relax
\mciteBstWouldAddEndPuncttrue
\mciteSetBstMidEndSepPunct{\mcitedefaultmidpunct}
{\mcitedefaultendpunct}{\mcitedefaultseppunct}\relax
\EndOfBibitem
\bibitem[Franosch \emph{et~al.}(2011)Franosch, Spanner, Bauer, Schr\"oder-Turk,
  and H\"of{}ling]{Franosch2011}
T.~Franosch, M.~Spanner, T.~Bauer, G.~E. Schr\"oder-Turk and F.~H\"of{}ling,
  \emph{J. Non-Cryst. Solids}, 2011, \textbf{357}, 472--478\relax
\mciteBstWouldAddEndPuncttrue
\mciteSetBstMidEndSepPunct{\mcitedefaultmidpunct}
{\mcitedefaultendpunct}{\mcitedefaultseppunct}\relax
\EndOfBibitem
\bibitem[Spanner \emph{et~al.}(2013)Spanner, Schnyder, H{\"o}f{}ling,
  Voigtmann, and Franosch]{Spanner2013}
M.~Spanner, S.~K. Schnyder, F.~H{\"o}f{}ling, T.~Voigtmann and T.~Franosch,
  \emph{Soft Matter}, 2013, \textbf{9}, 1604--1611\relax
\mciteBstWouldAddEndPuncttrue
\mciteSetBstMidEndSepPunct{\mcitedefaultmidpunct}
{\mcitedefaultendpunct}{\mcitedefaultseppunct}\relax
\EndOfBibitem
\bibitem[Chubynsky and Slater(2014)]{Chubynsky2014}
M.~V. Chubynsky and G.~W. Slater, \emph{Phys. Rev. Lett.}, 2014, \textbf{113},
  098302\relax
\mciteBstWouldAddEndPuncttrue
\mciteSetBstMidEndSepPunct{\mcitedefaultmidpunct}
{\mcitedefaultendpunct}{\mcitedefaultseppunct}\relax
\EndOfBibitem
\bibitem[Narayanan \emph{et~al.}(2006)Narayanan, Xiong, and Liu]{Narayanan2006}
J.~Narayanan, J.-Y. Xiong and X.-Y. Liu, J. Phys.: Conf. Series, 2006,
  p.~83\relax
\mciteBstWouldAddEndPuncttrue
\mciteSetBstMidEndSepPunct{\mcitedefaultmidpunct}
{\mcitedefaultendpunct}{\mcitedefaultseppunct}\relax
\EndOfBibitem
\bibitem[Pluen \emph{et~al.}(1999)Pluen, Netti, Jain, and Berk]{Pluen1999}
A.~Pluen, P.~A. Netti, R.~K. Jain and D.~A. Berk, \emph{Biophysical journal},
  1999, \textbf{77}, 542--552\relax
\mciteBstWouldAddEndPuncttrue
\mciteSetBstMidEndSepPunct{\mcitedefaultmidpunct}
{\mcitedefaultendpunct}{\mcitedefaultseppunct}\relax
\EndOfBibitem
\bibitem[Maaloum \emph{et~al.}(1998)Maaloum, Pernodet, and
  Tinland]{Maaloum1998}
M.~Maaloum, N.~Pernodet and B.~Tinland, \emph{Electrophoresis}, 1998,
  \textbf{19}, 1606--1610\relax
\mciteBstWouldAddEndPuncttrue
\mciteSetBstMidEndSepPunct{\mcitedefaultmidpunct}
{\mcitedefaultendpunct}{\mcitedefaultseppunct}\relax
\EndOfBibitem
\bibitem[Powles \emph{et~al.}(1992)Powles, Mallett, Rickayzen, and
  Evans]{Powles1992}
J.~G. Powles, M.~Mallett, G.~Rickayzen and W.~Evans, Proceedings of the Royal
  Society of London A: Mathematical, Physical and Engineering Sciences, 1992,
  pp. 391--403\relax
\mciteBstWouldAddEndPuncttrue
\mciteSetBstMidEndSepPunct{\mcitedefaultmidpunct}
{\mcitedefaultendpunct}{\mcitedefaultseppunct}\relax
\EndOfBibitem
\bibitem[Saxton(2007)]{Saxton2007}
M.~J. Saxton, \emph{Biophys. J.}, 2007, \textbf{92}, 1178--1191\relax
\mciteBstWouldAddEndPuncttrue
\mciteSetBstMidEndSepPunct{\mcitedefaultmidpunct}
{\mcitedefaultendpunct}{\mcitedefaultseppunct}\relax
\EndOfBibitem
\bibitem[Destainville \emph{et~al.}(2008)Destainville, Sauli{\`e}re, and
  Salom{\'e}]{Destainville2008}
N.~Destainville, A.~Sauli{\`e}re and L.~Salom{\'e}, \emph{Biophysical journal},
  2008, \textbf{95}, 3117\relax
\mciteBstWouldAddEndPuncttrue
\mciteSetBstMidEndSepPunct{\mcitedefaultmidpunct}
{\mcitedefaultendpunct}{\mcitedefaultseppunct}\relax
\EndOfBibitem
\bibitem[Murase \emph{et~al.}(2004)Murase, Fujiwara, Umemura, Suzuki, Iino,
  Yamashita, Saito, Murakoshi, Ritchie, and Kusumi]{Murase2004}
K.~Murase, T.~Fujiwara, Y.~Umemura, K.~Suzuki, R.~Iino, H.~Yamashita, M.~Saito,
  H.~Murakoshi, K.~Ritchie and A.~Kusumi, \emph{Biophysical journal}, 2004,
  \textbf{86}, 4075--4093\relax
\mciteBstWouldAddEndPuncttrue
\mciteSetBstMidEndSepPunct{\mcitedefaultmidpunct}
{\mcitedefaultendpunct}{\mcitedefaultseppunct}\relax
\EndOfBibitem
\bibitem[Fundueanu \emph{et~al.}(1999)Fundueanu, Nastruzzi, Carpov, Desbrieres,
  and Rinaudo]{Fundueanu1999}
G.~Fundueanu, C.~Nastruzzi, A.~Carpov, J.~Desbrieres and M.~Rinaudo,
  \emph{Biomaterials}, 1999, \textbf{20}, 1427--1435\relax
\mciteBstWouldAddEndPuncttrue
\mciteSetBstMidEndSepPunct{\mcitedefaultmidpunct}
{\mcitedefaultendpunct}{\mcitedefaultseppunct}\relax
\EndOfBibitem
\bibitem[Massignan \emph{et~al.}(2014)Massignan, Manzo, Torreno-Pina,
  García-Parajo, Lewenstein, and Lapeyre]{Massignan2014}
P.~Massignan, C.~Manzo, J.~A. Torreno-Pina, M.~F. García-Parajo, M.~Lewenstein
  and G.~J. Lapeyre, \emph{Phys. Rev. Lett.}, 2014, \textbf{112}, 150603\relax
\mciteBstWouldAddEndPuncttrue
\mciteSetBstMidEndSepPunct{\mcitedefaultmidpunct}
{\mcitedefaultendpunct}{\mcitedefaultseppunct}\relax
\EndOfBibitem
\bibitem[Manzo \emph{et~al.}(2015)Manzo, Torreno-Pina, Massignan, Lapeyre,
  Lewenstein, and Garcia~Parajo]{Manzo2015}
C.~Manzo, J.~A. Torreno-Pina, P.~Massignan, G.~J. Lapeyre, M.~Lewenstein and
  M.~F. Garcia~Parajo, \emph{Phys. Rev. X}, 2015, \textbf{5}, 011021\relax
\mciteBstWouldAddEndPuncttrue
\mciteSetBstMidEndSepPunct{\mcitedefaultmidpunct}
{\mcitedefaultendpunct}{\mcitedefaultseppunct}\relax
\EndOfBibitem
\bibitem[Vagias \emph{et~al.}(2013)Vagias, Raccis, Koynov, Jonas, Butt, Fytas,
  Ko{\v{s}}ovan, Lenz, and Holm]{Vagias2013}
A.~Vagias, R.~Raccis, K.~Koynov, U.~Jonas, H.-J. Butt, G.~Fytas,
  P.~Ko{\v{s}}ovan, O.~Lenz and C.~Holm, \emph{Phys. Rev. Lett.}, 2013,
  \textbf{111}, 088301\relax
\mciteBstWouldAddEndPuncttrue
\mciteSetBstMidEndSepPunct{\mcitedefaultmidpunct}
{\mcitedefaultendpunct}{\mcitedefaultseppunct}\relax
\EndOfBibitem
\bibitem[Porcher \emph{et~al.}(2010)Porcher, Abu-Arish, Huart, Roelens, Fradin,
  and Dostatni]{Porcher2010}
A.~Porcher, A.~Abu-Arish, S.~Huart, B.~Roelens, C.~Fradin and N.~Dostatni,
  \emph{Development}, 2010, \textbf{137}, 2795--2804\relax
\mciteBstWouldAddEndPuncttrue
\mciteSetBstMidEndSepPunct{\mcitedefaultmidpunct}
{\mcitedefaultendpunct}{\mcitedefaultseppunct}\relax
\EndOfBibitem
\bibitem[Saxton(2012)]{Saxton2012}
M.~J. Saxton, \emph{Biophys. J.}, 2012, \textbf{103}, 2411--2422\relax
\mciteBstWouldAddEndPuncttrue
\mciteSetBstMidEndSepPunct{\mcitedefaultmidpunct}
{\mcitedefaultendpunct}{\mcitedefaultseppunct}\relax
\EndOfBibitem
\end{mcitethebibliography}

\providecommand*{\mcitethebibliography}{\thebibliography}
\csname @ifundefined\endcsname{endmcitethebibliography}
{\let\endmcitethebibliography\endthebibliography}{}

}

\end{document}